\documentclass[useAMS]{mn2e}
\newcommand{\kms}{{\rm {km\, s^{-1}}}}

\newcommand{\K} {\rm K}

\newcommand{\etal}{{et al.}~}

\def\dd{\,{\rm d}}
\def\ltsima{$\; \buildrel < \over \sim \;$}
\def\lsim{\lower.5ex\hbox{\ltsima}}
\def\gtsima{$\; \buildrel > \over \sim \;$}
\def\gsim{\lower.5ex\hbox{\gtsima}}
\def\ga{\mathrel{\hbox{\rlap{\hbox{\lower4pt\hbox{$\sim$}}}\hbox{$>$}}}}
\def\la{\mathrel{\hbox{\rlap{\hbox{\lower4pt\hbox{$\sim$}}}\hbox{$<$}}}}

\newcommand{\num}{\nu_{\rm c}}
\newcommand{\ttau}{\tilde \tau}
\newcommand{\ttaum}{\ttau_{\rm c}}
\newcommand{\ttaut}{\ttau_{\rm t}}

\def\vp{v_{\rm p}}

\def\cA{${\cal A}$}

\newcommand{\bc}{\begin{center}}
\newcommand{\be}{\begin{equation}}
\newcommand{\ee}{\end{equation}}
\newcommand{\ec}{\end{center}}

\newcommand{\hi}{\mbox{H{\scriptsize I}}}

\newcommand{\dhi}{n_{\mathrm{HI}}}
\newcommand{\dhhi}{\hat n_{\mathrm{HI}}}
\newcommand{\vx}{{\bf x}}

\newcommand{\Mpc}{\rm Mpc}

\newcommand{\lya}{\mbox{Ly$\alpha$~}}
\newcommand{\eg}{{e.g.,~}}
\newcommand{\ie}{{i.e.,~}}

\newcommand{\hep} {{\rm He{\sc ii}}}
\newcommand{\hepp} {{\rm He{\sc iii}}}

\def\nhi{\dhi}

\def \simlt{\lower.5ex\hbox{\ltsima}}            
\def \gtsima{\mbox{$\; \buildrel > \over \sim \;$}}
\def \simgt{\lower.5ex\hbox{\gtsima}}            

\newcommand{\mnras}{MNRAS}
\newcommand{\apj}{ApJ}

\newcommand{\aap}{A\&A}

\title[The Matter Power Spectrum From The \lya Forest] 
{The matter power spectrum from the
\lya forest: an optical depth estimate}

\author[Zaroubi \etal]{S. Zaroubi$^{1}$, M. Viel$^{2}$,
A. Nusser$^{3,4}$, M. Haehnelt$^{2}$ and T.-S. Kim$^{2}$\\ $^{\,1}$
Kapteyn Astronomical Institute, University of Groningen, P.O. Box 800,
9700 AV Groningen, the Netherlands\\ $^{\,2}$ Institute of Astronomy,
Madingley Road, Cambridge CB3 0HA \\ $^{\,3}$ Physics Department,
Technion--Israel Institute of Technology, Technion City, Haifa 32000,
Israel\\ $^{\,4}$Division of Theoretical Astrophysics National
Astronomical Observatory Japan, Mitaka 181-8588, Japan}

\begin{document}

\maketitle

\begin{abstract} 
We measure the matter power spectrum from 31 \lya
spectra spanning the redshift range of 1.6-3.6. The optical depth,
$\tau$, for \lya absorption of the intergalactic medium is obtained
from the flux using the inversion method of Nusser \& Haehnelt (1999).
The optical depth is converted to density by using a simple power law relation,
$\tau \propto (1+\delta)^\alpha$.  The non-linear 1D power spectrum
of the gas density is then inferred with a method that makes
simultaneous use of the 1 and 2 point statistics of the flux 
and compared against theoretical models with a likelihood analysis. 
A Cold Dark Matter (CDM) model with standard cosmological parameters fits
the data well.  The power spectrum amplitude is measured to be
(assuming a flat Universe), $\sigma_8 = (0.9 \pm 0.09)\times
(\Omega_{\rm m}/0.3)^{-0.3}$, with $\alpha$ varying in the range of
$1.56-1.8$ with redshift. Enforcing the same cosmological parameters
in all four redshift bins, the likelihood analysis suggests 
some evolution in the density-temperature
relation and the thermal smoothing length of the gas. The inferred
evolution is consistent with that expected if reionization of \hep\
occurred at $z\sim 3.2$.  A joint analysis with the WMAP results 
together with  a prior on the Hubble constant as suggested by the HST
key project data, yields values of $\Omega_{\rm m}$
and $\sigma_8$ that are consistent with the cosmological concordance
model. We also perform a further inversion to obtain the linear 3D
power spectrum of the matter density fluctuations.
\end{abstract}


\section{Introduction}

The numerous \lya absorption features observed in quasar spectra
blue-wards of their \lya emission line known as the \lya forest,
provide one of the main probes of the intergalactic medium (hereafter
IGM) (Bahcall \& Salpeter 1965; Gunn \& Peterson 1965).  In recent
years two main advances have shaped the accepted view on the origin of
the \lya absorbing structures. First, the advent of 10-meter class
telescopes equipped with high-resolution echelle spectrographs (HIRES
on Keck and UVES on the Very Large Telescope) has provided us with
data of unprecedented quality (see Rauch 1998 for a review).  Second,
the emergence of a theoretical paradigm within the context of the cold
dark matter (CDM) cosmology (e.g. Bi, Boerner \& Chu 1992) supported
by numerical hydrodynamical simulations (Cen \etal 1994; Zhang,
Anninos \& Norman 1995; Miralda-Escude´ \etal 1996; Hernquist \etal
1996; Wadsley \& Bond 1996; Zhang \etal 1997; Theuns \etal 1998;
Machacek \etal 2000; Viel, Haehnelt \& Springel 2004; Tytler et
al. 2004; Jena et al. 2005) and semi-analytical studies (e.g. Pichon
\etal 2001; Matarrese \& Mohayaee 2002; Viel \etal 2002). According to
this paradigm, the absorption is produced by volume filling
photoionized gas that contains most of the baryons at redshifts
$z\approx 3$ (see e.g. Efstathiou, Schaye \& Theuns (2000) for a
recent review), where the absorbers are locally overdense extended
structures, close to local hydrostatic equilibrium (Schaye 2001). The
paradigm also predicts that most of the gas probed by the \lya forest
- absorption features with column density $\lsim 10^{13.5} {\rm
cm}^{-2}$ - resides in mildly non-linear dark-matter overdensities.

On scales smaller than the Jeans scale the baryonic gas 
is smoothed by pressure forces, erasing the small-scale fluctuations
of the gas density, and setting its distribution apart from the
 dark matter component. On these small scales the width of the absorption
features is determined by the gas thermodynamical properties enabling 
the measurement of the IGM temperature and temperature-density
relation (\eg\ Schaye \etal 2000, Theuns \& Zaroubi 2000, McDonald \etal 2001, 
Theuns \etal 2002a and 2002b, Gleser \etal 2005).  

On scales larger than the Jeans scale, however, the gas distribution
faithfully follows that of the underlying dark matter. The gas
distribution on these scales provides a probe of the dark matter
distribution and its power spectrum (Croft \etal 1998;
Hui 1999; Croft \etal 2002, McDonald \etal 2000, Hui \etal 2001; Croft
\etal 2002; Viel \etal 2003, 2004, 2005; McDonald et al. 2005; Seljak
et al. 2005; Desjacques \& Nusser 2005; Lidz et al. 2005, Viel \&
Haehnelt 2005).

The ``standard'' method used for measuring the matter power spectrum
from the \lya forest is based on the result obtained in numerical
simulations showing that the normalized flux power spectrum is
proportional to that of the underlying matter (Croft \etal 1998). The
calibration of the relation between the two relies  on
numerical simulations and on the value of the mean flux. Both of these
are somewhat uncertain.  The calibration of the relation depends on the specific
cosmological parameters of the simulation, and the mean flux on the
specific quasar (QSO) spectrum at hand.

To date, almost all matter power spectra inferred from the \lya forest
are based on this relation.  There is, however, some debate on what to
adopt for the mean level of the flux. Early power spectrum
measurements from the \lya forest (\eg\ Croft \etal 1998; 2002;
McDonald \etal 2000) adopted a rather low level of the mean flux, 
and inferred a relatively low amplitude of the fluctuations. This result is
not completely consistent with the large scale angular power spectrum
amplitude and the  early reionization of the Universe inferred
from the temperature and polarisation data of the WMAP
satellite (Kogut \etal 2003). This tension has been the primary reason
for the WMAP team to suggest a running spectral index model 
(Spergel \etal 2003).  A number of authors have recently pointed to
the strong dependence of the inferred amplitude on  the adopted
mean flux level and have argued that the errors 
of dark matter power spectrum inferred from \lya forest data 
have been underestimated (Zaldarriaga, Scoccimarro \& Hui, 2003; Zaldarriaga, Hui \& Tegmark,
2001; Gnedin \& Hamilton 2002; Seljak, McDonald \& Makarov 2003).

Recent studies (Kim \etal 2004, Viel \etal 2004 and Seljak \etal 2005,
Lidz et al. 2005, Viel \& Haehnelt 2005) have all adopted the higher
values of the mean flux suggested by high-quality absorption
spectra. These studies obtained matter power spectra from \lya
data that are consistent with the WMAP results without the need for
the running spectral index power spectrum proposed by the WMAP team.

The current study follows a different route in which the matter power
spectrum is measured by inverting the normalized \lya flux to obtain
the optical depth for \lya absorption (Nusser \& Haehnelt 1999, 2000;
hereafter NH99 and NH00 respectively). In a system in
photoionization-recombination equilibrium, such as the IGM, the
optical depth for \lya absorption is to a good approximation a power
law function of the underlying density with the power law index
determined by the temperature-density relation of the IGM gas.  
We use this  simple
relation to infer the line-of-sight distribution of the gas
density. The shape of the power spectrum and the probability
distribution of the gas density, are then used to infer the shape and
amplitude of the 1D-power spectrum of the gas density separately,
without the need for assuming a mean flux level.  State-of-the-art
hydrodynamical simulations are then used  to calibrate and test the
method.  The method is applied to 31 high
resolution \lya spectra of which the newly acquired LUQAS sample
constitute the main part.

The paper is organised as follow: \S 2 describes the data set.  \S 3
shows how we infer the non-linear 1D power spectrum of the gas density
from the \lya forest data using the NH99 \& NH00 inversion method. \S
4 presents the analytical modelling of the non-linear 1D power
spectrum of the gas from the linear 3D power spectrum of matter
and describes the use of numerical hydro-simulations to correct for
the effect of redshift distortions and gas pressure.  The likelihood
analysis employed to constrain the parameters of the matter power
spectrum and the thermal state of the IGM, is discussed in \S 5. In \S
6, our  best estimate of the 3D matter power spectrum is presented and
compared to results from previous studies.  The main conclusions are
given in \S 7.

\section{The data set}

The sample used in this study consists of 31 high-resolution high S/N
spectra. 27 of the spectra were obtained with the Ultra-Violet Echelle
Spectrograph (UVES) on VLT, Paranal, Chile, over the period
1999-2002. The 27 spectra were taken from the ESO archive and are
publicly available (P.I.: J. Bergeron [Bergeron \etal 2004]); this
sample is known as the LUQAS sample (Kim \etal 2004). The LUQAS sample
was selected based on the following criteria: 1) S/N larger than 25 in
the Lyman-$\alpha$ forest region; 2) complete or nearly complete
coverage of the Lyman-$\alpha$ forest region; 3) no damped-\lya (DLA)
systems in the \lya forest region, though few spectra have sub-DLAs
(column density $10^{19.0-20.3}$ cm$^{-2}$); 4) no broad absorption
line systems; 5) publicly available as of January 1, 2003.  The total
redshift path of the sample is z = 13.75. For more details on the
LUQAS sample see Kim \etal (2004).  The rest of the spectra were
obtained from various other publicly available spectra that fulfill
similar criteria.  For more details on the remaining spectra, see Hu
\etal (1995) and Theuns \etal (2002a) and references therein. All the
31 spectra used here, have signal-to-noise ratios of 40-50 per pixel,
and a similar resolution, ($\lambda/\Delta\lambda \gsim 40000$).

\begin{figure}
\setlength{\unitlength}{1cm} \centering
\begin{picture}(8,11)
\put(-2.0,-2.8){\includegraphics{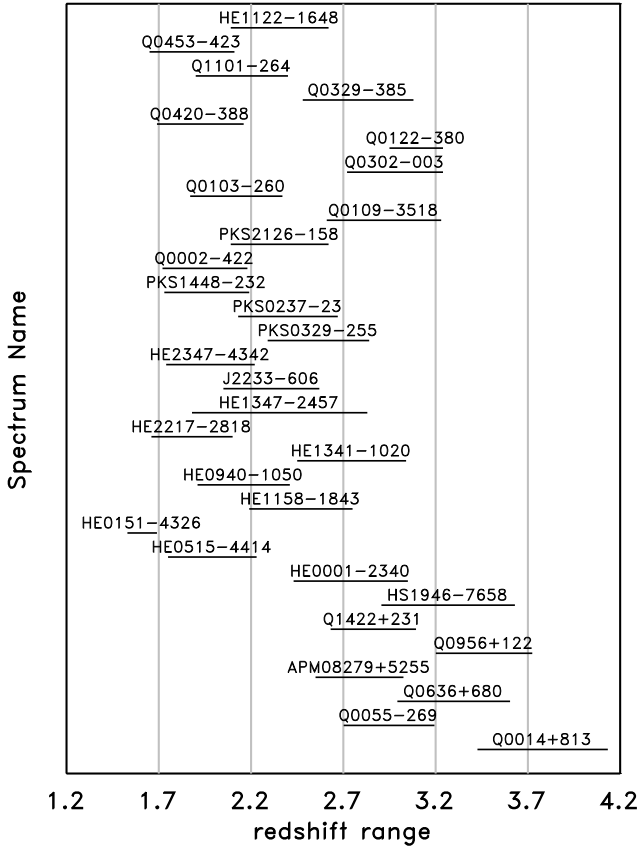}}
\end{picture}
\caption{The redshift range of the QSO spectra used in this paper. 27
of the spectra are taken from the LUQAS QSO sample (Bergeron \etal
2004; Kim \etal 2004). QSO 0956+122 is taken from Hu \etal (1995). The
remaining 4, at the high redshift end of the sample, are various other
publicly available spectra  (see Theuns \etal (2002a) and references
therein). }
\label{fig:zrange}
\end{figure}

Figure~\ref{fig:zrange} shows the redshift range covered by all the
spectra used in this analysis. The median redshift of the sample is
$\langle z \rangle = 2.55$ with a cumulative redshift path of about
16.8.

\section{Recovering the 1D power spectrum of the gas density in redshift space}
\label{sec:1D-gas}

\subsection{From flux to gas density}

\label{sec:flux-gas}

Nusser \& Haehnelt (NH99 \& NH00) introduced an algorithm to invert
the observed flux in the \lya forest region of QSO absorption spectra
to obtain the gas density along the line-of-sight (see also Pichon
\etal 2001).  We will use here a
modified version of this algorithm to measure the 1D power spectrum of
the gas density. This is the first step in our endeavour to constrain
the matter power spectrum and the thermal state of the IGM.  For the
sake of completeness, the description of the algorithm is repeated
here in some detail.

The optical depth in redshift space due to resonant \lya scattering 
is related to the \hi\ density along the line of sight (LOS) in real space by 
\begin{equation}
\tau(z)=  \sigma_{0} \;\frac{c}{H(z)}\;\int_{-\infty}^{\infty} \dhi (z, x) 
{\cal H}[z-x-\vp(x) , b(x)]  
\dd x,  \\
\nonumber
\label{tau}
\end{equation}
where $\sigma_{0}$ is the effective cross section for resonant line
scattering, $H(z)$ is the Hubble constant at redshift $z$, $x$ is the 
real space coordinate (in $\kms$), {$\cal H$} is the Voigt profile 
normalized such that $\int{\cal H }\, \dd x =1 $, $v_{\rm p}(x)$ is 
the LOS peculiar velocity, and $b(x)$ is the Doppler parameter due to 
thermal/turbulent broadening.  The absorption features in the \lya\
forest are mainly produced by regions of low to moderate densities, where
photoheating is the dominant heating source and shock heating  is not
important.  

Hydrogen in the IGM is highly ionized (Gunn \& Peterson 1965, Scheuer
1965) and the photoionization equilibrium in the expanding IGM
establishes a tight correlation between neutral and total hydrogen
density.  Numerical simulations have supported the existence of this
correlation and shown that the gas density traces the fluctuations of
the DM density on scales larger than the Jeans length, so that $ \dhi
= \dhhi \left(\frac{\rho_{\mathrm{DM}}(\vx)}{ \bar
\rho_{\mathrm{DM}}}\right)^\alpha$ . Here $\dhhi$ is the \hi\ density
at the mean dark matter density, and the parameter $\alpha$ depends on
the reionization history. The possible range for $\alpha$ is $1.56 \la
\alpha \la 2$ with a value close to $2\;$ just after reionization, and
decreasing at later times (Hui \& Gnedin 1997). In this relation ${\rho_{_{\rm
DM}}(\vx)}$  is the dark matter density smoothed on
the Jeans length below which thermal pressure becomes important.
The Jeans length in comoving units in the linear regime is given by,
\begin{eqnarray} 
X_{_{\rm J}} &=& \frac {2\pi c_{\rm s}}{\sqrt{4\pi G \bar \rho}} \,
(1+z) \nonumber \\ 
& \approx & 1.3 
\left ( \frac{\Omega_{\rm m} h^2}{0.125} \right )^{-1/2} 
\left ( \frac{\bar T}{1.5 \times 10^4\K} \right )^{1/2} \times   \nonumber\\
& & \left ( \frac{1.5}{1 + (2-\alpha)/0.7} \right )^{1/2} 
\left ( \frac{1+z}{4} \right )^{-1/2} \; \Mpc,
\end{eqnarray} 
where $c_{\rm s}$ is the sound speed, $\bar \rho$ is the mean density
of the Universe, $\Omega_{\rm m}$ is the matter density parameter,
$\bar T$ is the mean IGM temperature and $h$ is the Hubble constant of
units of $100 \mathrm{\kms \Mpc^{-1}}$.  However, in the non-linear
regime gas can collapse to scales smaller than this and the Jeans
scale becomes a somewhat ambiguous quantity.  The effective non-linear
Jeans length, $X_{_{\rm J}}$, is defined as the width of a kernel of
the form $[1+ (k X_{_{\rm J}}/2\pi)^2]^{-2}$, such that the rms
fluctuation amplitude of $ {\rho_{_{\rm DM}}(\vx)}$ is the same as
that of the unsmoothed dark matter density filtered with this kernel
(see section 4 for details).  On scales larger than the effective
Jeans length, equation (\ref{tau}) can be written as
\begin{equation}
\tau(z, w)=  {\cal A} (z)  \int_{-\infty}^{\infty}  
\left(\frac{\rho_{_{\rm DM}}(z, \vx)}{{\bar \rho_{_{\rm DM}}}}\right)^\alpha
{\cal H} [w-x-\vp (x), b(x)] \dd x ,
\label{tauw}
\end{equation}
with
\begin{eqnarray}
{\cal A} (z)&=& \sigma_{0}  \frac{c}{H(z)} \; \dhhi  \nonumber \\
&\approx&   0.61 \; 
\left (\frac{300\, \kms \Mpc^{-1}}{H(z)}\right ) \;
\left (\frac{\Omega_{\rm b}h^2}{0.02}\right )^{2} \times   \nonumber\\ 
&&\left ( \frac{\Gamma_{\rm phot}}{10^{-12}\,s^{-1}} \right)^{-1}\; 
\left ( \frac{\bar T}{1.5\times 10^{4}\K} \right )^{-0.7} \;
\left ( \frac{1+z}{4} \right )^{6} , \nonumber \\
\label{tauf}
\end{eqnarray}
where $\Omega_{\rm b}$ is the baryonic density in terms of the
critical density, $H(z)$ is the Hubble parameter, 
and $\Gamma_{\rm phot}$ is the photoionization rate per
hydrogen atom. The Doppler parameter in the last equation depends on
$\nhi$ as $b\propto \dhi^{1-\alpha/2}$.

NH99 defined the  local optical depth as,   
\begin{equation}
\tilde\tau(x)\equiv {\cal A} \left[\frac{\rho(x)}{{\bar \rho}}\right]^{\alpha} ,
\label{eq:lod}
\end{equation} 
which is related to the 
observed optical depth $\tau$ by a convolution with a Voigt
profile as described in equation  (\ref{tau}). 

NH99 have presented a direct Lucy-type iterative scheme (Lucy 1970) to
recover the optical depth and the corresponding mass and velocity
fields in the LOS from the normalized flux, $F=\exp(-\tau)$. In our 
tests with hydro-simulation (described in more detail in
\S~\ref{sec:numerical}), we found that estimating the velocity field
from the spectra itself is not very accurate. We will therefore be
less ambitious here and use the algorithm of NH99 to recover the gas
density in redshift space. We will later use hydrodynamical
simulations to address the effect of redshift space distortions (see
\S~\ref{sec:distortions})

NH99 showed that the density field can be successfully recovered below
a threshold value above which the corresponding flux saturates.  The
NH99 reconstruction method, therefore, imposes an effective upper limit
on the recovered optical depth in these regions.  This will inevitably
affect the amplitude of the measured power spectrum but as will be
shown later, not its shape.  We use this to determine the shape and 
normalization of the power spectrum separately in two steps.

\begin{figure*}
\setlength{\unitlength}{1cm} \centering
\begin{picture}(17,9)
\put(-2.5, -3.){\includegraphics{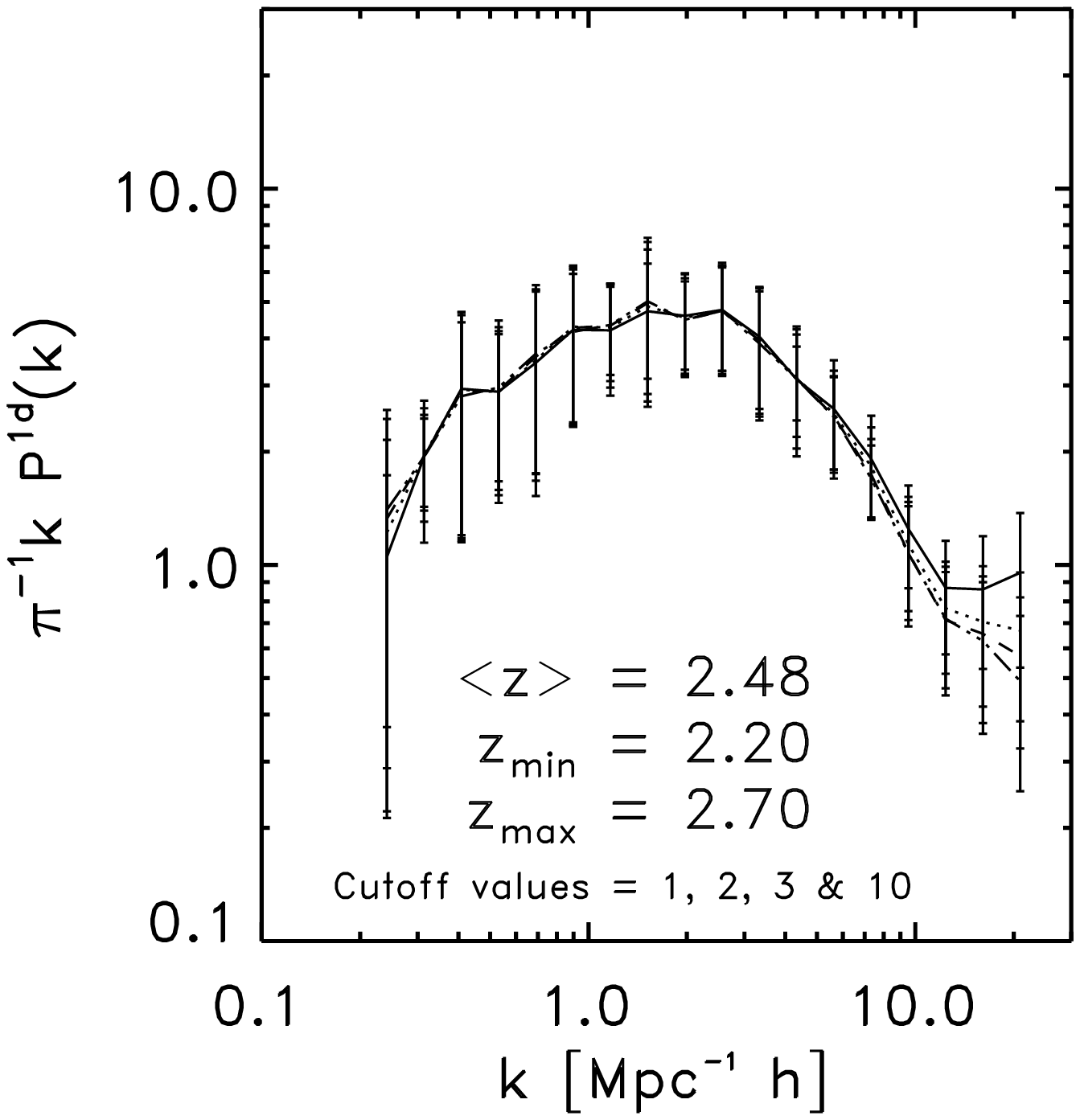}}
\put(7.2, -3.){\includegraphics{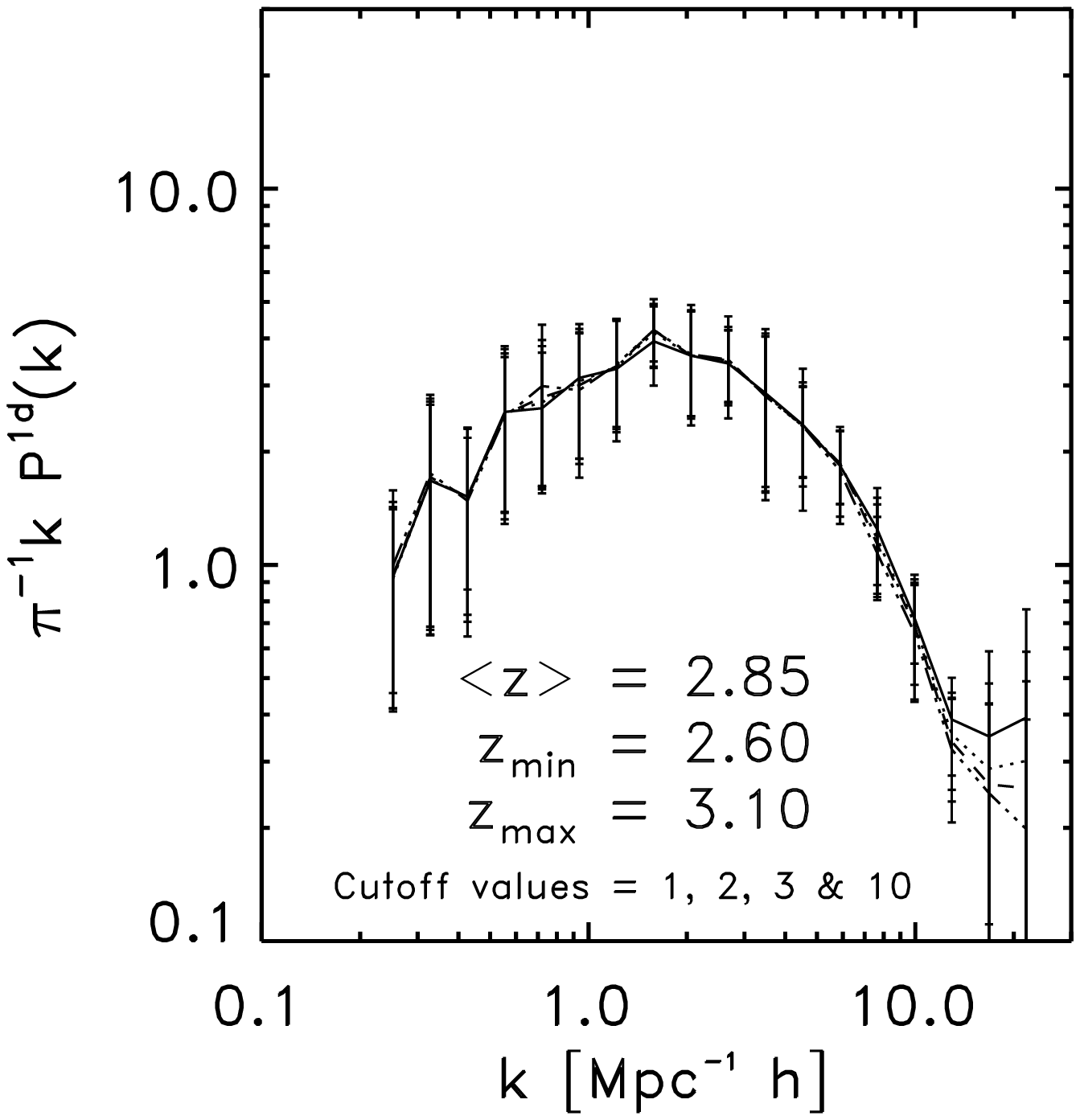}}
\end{picture}
\caption{The measured shape of the 1D power spectrum of the gas
density at two different redshifts for a range of 
$\ttau$-cutoff values.  The four cutoff values are $\ttau_\mathrm{cutoff} =$
1, 2, 3 and 10 and the spectra are normalized to the same amplitude.
The power spectra were calculated from all sections of the absorption
spectra in the specified redshift bin. The error bars show the
$1-\sigma$ errors. For $k\lsim 10\mathrm{Mpc^{-1}h}$ where the \lya
forest is dominant, the changes in the shape of the power spectrum
shape are very small. At small scales with $k \gsim
20\mathrm{Mpc^{-1}h}$, however, associated metal absorption results in
a clear change in the slope when the cutoff value is varied. Note that
only wavenumbers with values $\lsim 10\mathrm{Mpc^{-1}h}$ are used 
to estimate the 3D matter power spectrum.}
\label{fig:taucut}
\end{figure*}

\subsection{The shape of the 1D power spectrum of the gas density}
\label{sec:shape}

\label{obs_1D_PS}
Due to the 1-dimensional nature of the \lya forest, the measured power
spectrum at a given wave-number, $k$, is 
necessarily the line of sight power spectrum, $P^{1D}(k)$. The
relation between the 1D power spectrum and the three dimensional 
power spectrum, $P^{3D}(k)$ is (Kaiser \& Peacock 1991),

\begin{equation}
P^{3\mathrm{D}}(k) = -\frac{2 \pi}{k} \frac{\mathrm{d}P^{1\mathrm{D}}}{\mathrm{d}k}.
\label{eq:ps-diff}
\end{equation}

The useful dynamical range covered by the \lya forest data is restricted to $
0.1 \Mpc^{-1}\mathrm{h} \lsim k \lsim 10 \Mpc^{-1}\mathrm{h}$.  The
lower limit comes from the limited length of the observed QSO spectra.
The upper limit is imposed by a combination of the effective
Jeans scales below which the pressure gradients wipe out the small
scales fluctuations in the baryons and the contamination induced by metal
lines. In order to measure the 1D power spectrum from the 31 QSO
spectra, we have divided the \lya data into four redshift bins (see
table~\ref{table} and figure~\ref{fig:zrange}). For each redshift bin,
the sections that are taken into account are those that belong to \lya
spectra that have more than 40\% of their total length in 
the redshift bin at hand (see table~\ref{table}).  In order to measure
the power spectrum within a certain redshift bin, we first calculate
the local optical depth for each section within this bin using 
the NH99 method.  For a given value of $\alpha$, the 1D matter density
is  calculated from equation~\ref{eq:lod}.  The recovered density
section is then Fourier-transformed and the power spectrum at a given
wavenumber is obtained.  We have estimated the mean and the measurement error of
the 1D power spectrum at a given wavenumber from all the
individual spectra within the bin. The variance is calculated using
two independent methods that give very similar results. The first is
a standard deviation measurement of all the power spectrum values at a
given wavenumber while the second  uses a bootstrap technique.

\begin{table}
\caption{A table describing the redshift bins used in this paper.
Columns number 1,2,3 and 4 show the redshift bin number; its mean,
minimum and maximum redshifts, respectively. Columns 5 and 6 show the
number of spectra included in each bin and its total length in $\kms$.}
\label{table}
\begin{tabular}{cccccc}
\hline
z bin & $\langle \mathrm{z} \rangle$ & z$_{\rm min}$ & z$_{\rm max}$ & \# of
Spectra & Length in ${\rm km/s}$\\
\hline
1 & 3.29 & 3.0  & 3.6 & 7  & 202658\\
2 & 2.85 & 2.6  & 3.1 & 10 & 308166\\
3 & 2.48 & 2.2  & 2.7 & 10 & 308002\\
4 & 1.96 & 1.6  & 2.2 & 13 & 486124\\
\hline
\end{tabular}
\end{table}

The recovered optical depth $\tilde \tau$ is a good approximation to
the true field only in regions with $\tilde \tau $ smaller than a certain
value.  For large optical depths, the recovered $\tilde \tau$
typically underestimates the true field.  As in NH00, we define a
truncated local optical depth $\ttaut$ as $\ttaut=\ttau $ for $
\ttau<\ttaum$, and $\ttaut=0$ otherwise.  In order to test how this
cut-off in the optical depth affects the shape of the 1D power
spectrum, the power spectrum inferred from a truncated optical depth
distribution is measured for a range of cut off values. The resulting
dimensionless 1D power spectra\footnote{We will generally use the
dimensionless 1D and 3D power spectra to express our results. These
are defined as $\pi^{-1} k P^{1D}$ for the 1D case, and
$\Delta^2\equiv (2 \pi^2)^{-1} k^3 P^{3D}$ for the 3D case.} are shown
in Figure~\ref{fig:taucut}, all renormalized to the same fiducial
amplitude. The left panel is for $\langle z \rangle =2.48$ and the
right panel is for $\langle z \rangle = 2.85$.  The error bars show
the variance around the mean at each point.  Notice, that these errors
are independent.  Changing the cut-off value of the optical depth has
very little effect at $k \lsim 10 \Mpc^{-1} h$ . This is perhaps not
too surprising as the regions in the spectrum where the flux is
saturated have a small volume filling factor.

At small scales with $k \gsim 10 \Mpc^{-1} h$ metal absorption lines
within the \lya-forest contaminate the signal ({\it e.g.} Kim et
al. 2004). We therefore truncate  the recovered 1D power spectrum 
at this wavenumber.

\begin{figure*}
\setlength{\unitlength}{1cm} \centering
\begin{picture}(9,19)
\put(-5.5, 7.5){\includegraphics{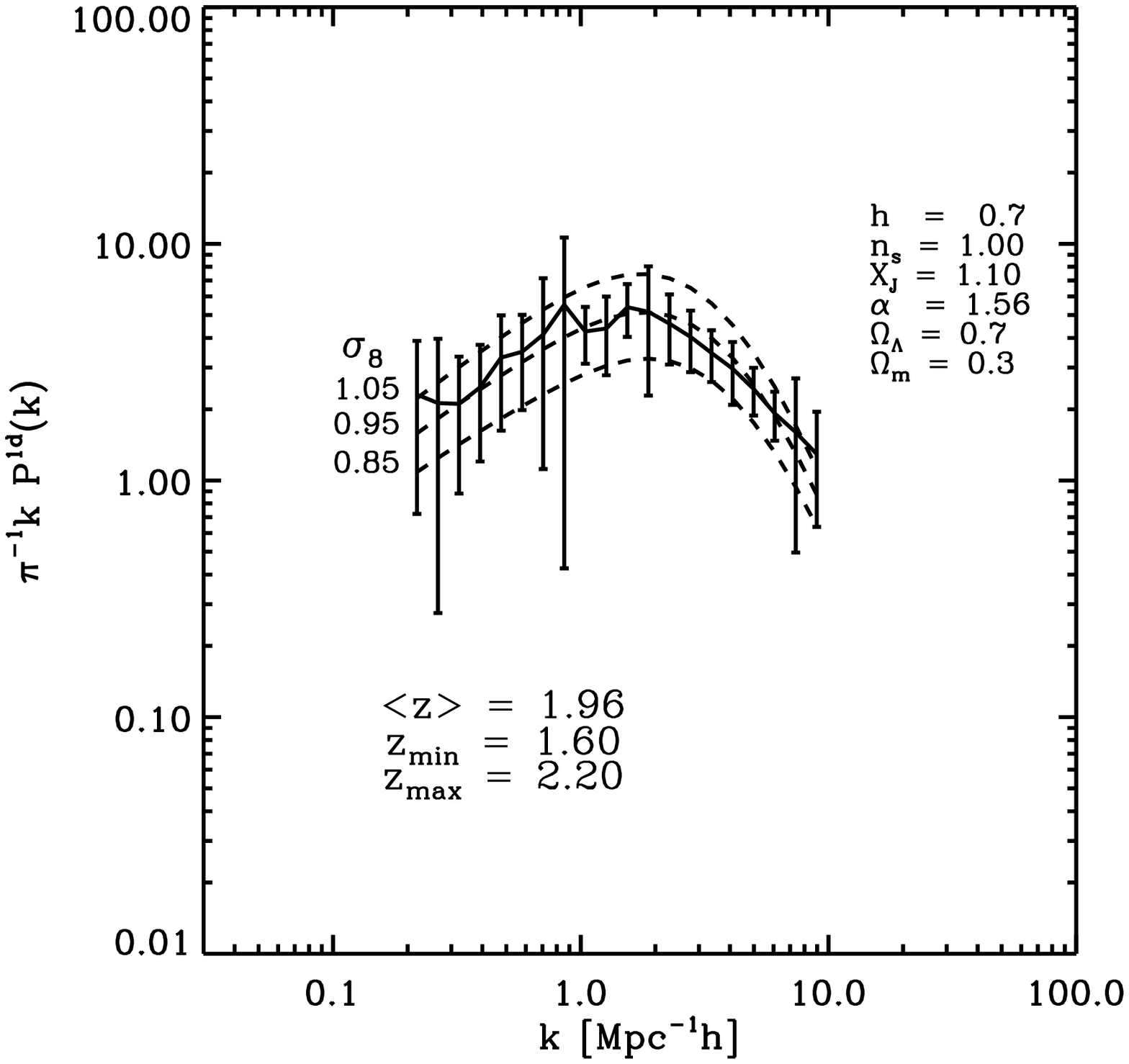}}
\put(4, 7.5){\includegraphics{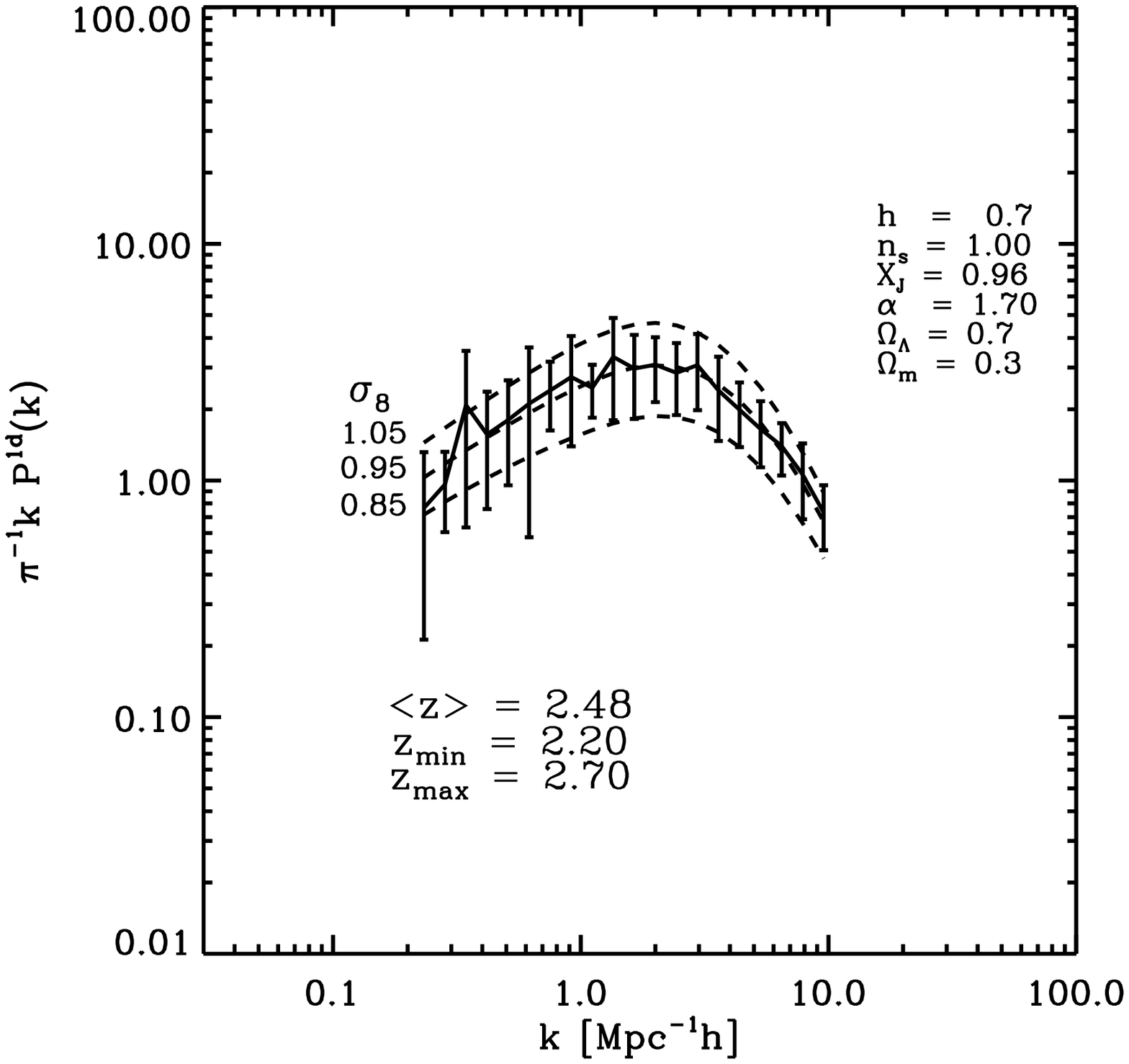}}
\put(-5.5, -2.){\includegraphics{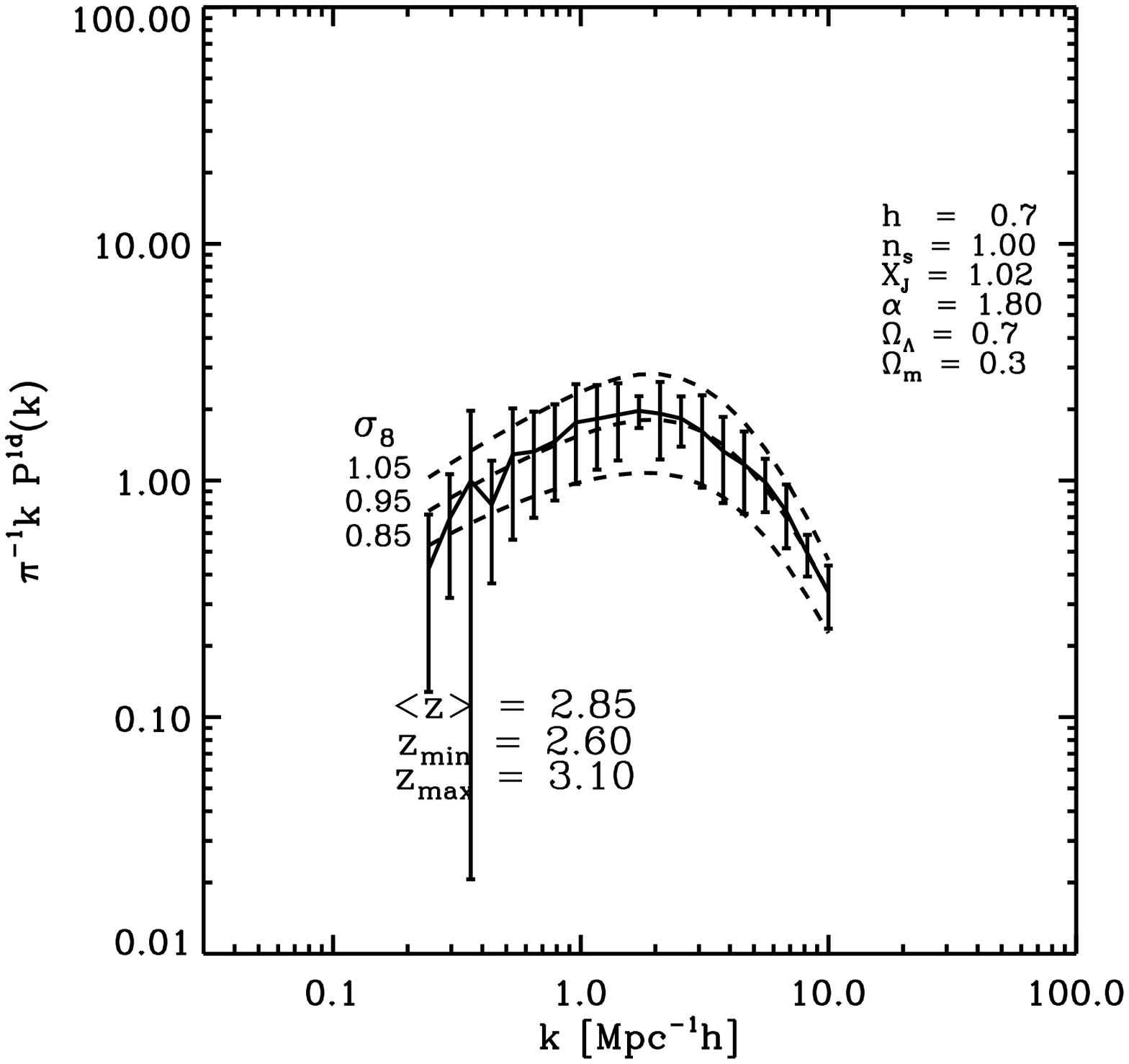}}
\put(4, -2.){\includegraphics{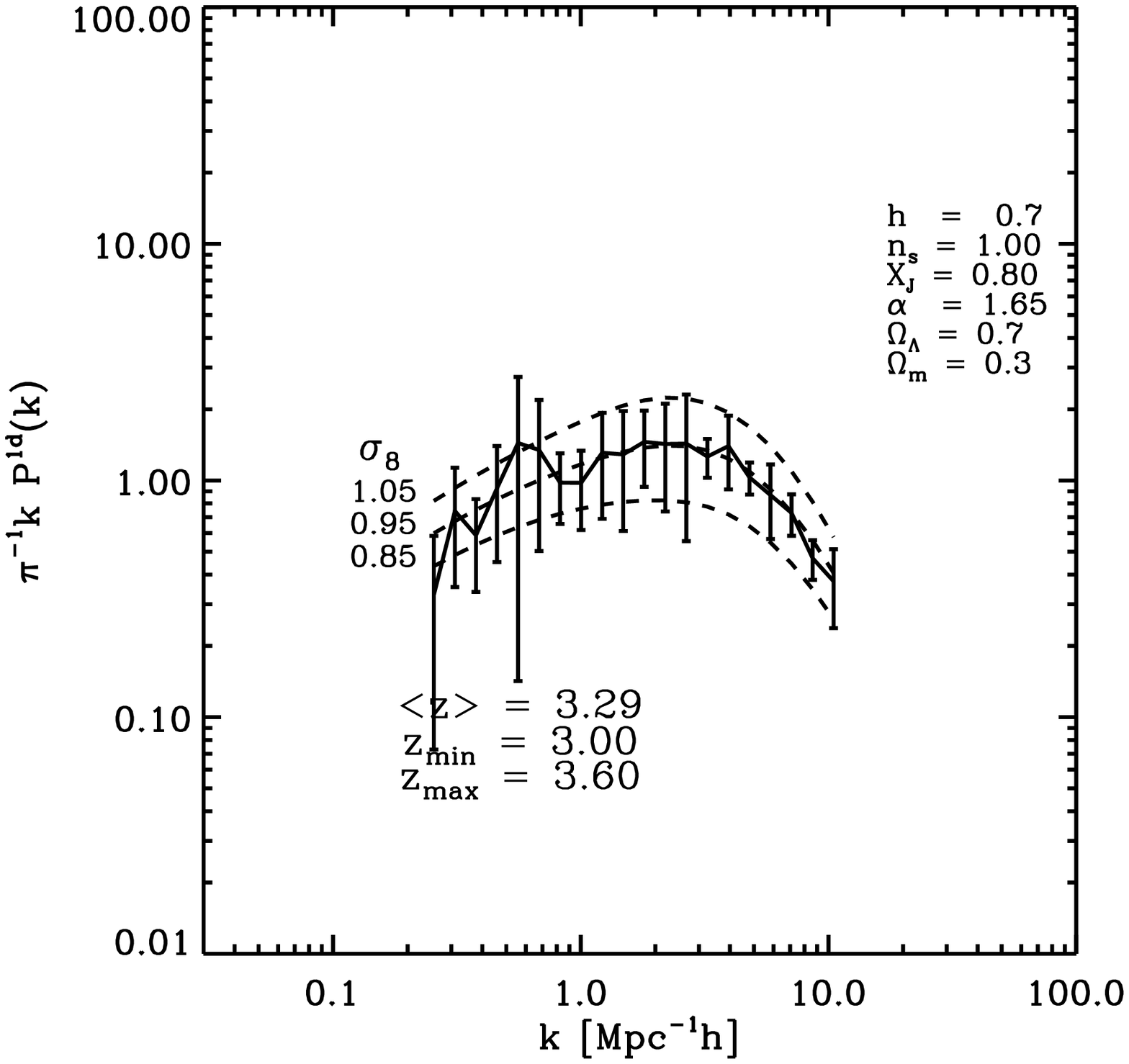}}
\end{picture}
\caption{The solid curves shows the measured 1D  power
spectra of the gas density for four different redshifts 
assuming the $\alpha$ values indicated on each panel. 
The error bars  include the combined contributions from the measurement of
shape and amplitude of the power spectrum to the
errors as described in the text. The central dashed curve in each
panel shows the best fitting models obtained from the likelihood
analysis. The other two dashed curves show the $\sim 1-\sigma$ range of 
allowed values of  $\sigma_8$. The parameters of the best fit models
are annotated on each panel. 
}
\label{fig:PS1D}
\end{figure*}

\subsection{The amplitude of the 1D power spectrum of the gas density}
\label{sec:amplitude}

Most of the regions where the \lya flux is saturated correspond to scales
comparable to the Jeans scale. These regions are scattered roughly randomly
across the spectrum. The amplitude of the 1D power spectrum of
the gas density can thus not be reliably measured directly from the optical
depth truncated at a certain cut-off value.

We will use instead the probability density function (PDF) of the
recovered optical depth which is related to the PDF of the gas
density. We thereby use the fact that the shape of the PDF changes
with increasing amplitude of the power spectrum.  NH00 showed that the
first two moments of the PDF of the gas density can be recovered rather
well from the moments of the probability distribution of the truncated
optical depth.  The main points are summarized in the following.

NH00 define the moments of the truncated  optical depth 
$\ttau_{\rm t}$ which can be written in terms of $\cal P$, the
density PDF,   as,
\begin{equation}
<\ttau^n_{\rm t}>= 
{\cal A}^n \int_{-\infty}^{\delta_{\rm c}} 
\left(1+\delta\right)^{n \alpha} {\cal P}(\delta) \dd \delta,
\label{trunc}
\end{equation}
where $\delta=\rho/{\bar \rho} -1$ is the density contrast and
$\delta_{\rm c}=\left(\ttaum/{\cal A}\right)^{1/\alpha}-1$.  Further,
$\nu$ is defined as, $\nu = [{\rm ln}(\rho/\bar \rho)-\mu_1]/\mu_2$,
where $\mu_1$ and $\mu_2$ are the average and $rms$ values of ${\rm
ln}\left(1+\delta\right)$.  NH99 had shown that the PDF of the DM
density smoothed on the scale relevant for the \lya forest (the
effective Jeans scale) can be reasonably well approximated by a
log-normal distribution (e.g. Bi \& Davidsen 1997 or Sheth 1998 and
Gaztanaga \& Croft 1999 for different forms of the PDF).  For a
log-normal density distribution, ${\cal P}(\nu)=\exp(-\nu^2/2)/\sqrt{2
\pi}$, the truncated moments in (\ref{trunc}) can be written as,
\begin{equation}
<\ttau^n_{\rm t}>=
\frac{{\cal A}^n}{2} \exp\left(\frac{1}{2}n^2 \alpha^2 \mu_2^2 + n \alpha \mu_1\right)
\left[1+ {\rm erf}\left(\frac{\num - n \alpha \mu_2}{\sqrt{2}} \right) \right]
\label{taumom}
\end{equation}
where $\num$ is the value of $\nu$ corresponding to $\delta_{\rm c}$. 
By  expressing $\nu$ in terms of $\ttau$ in (\ref{taumom}) 
 the truncated moments can be written as 
\begin{eqnarray}
<\ttau^n_{\rm t}> & = &
\frac{1}{2} \exp\left(\frac{1}{2}n^2 \alpha^2 \mu_2^2 - n \alpha \mu_2^2/2
+ n\ln {\cal A}    \right) \times   \nonumber\\
 & & \left[1+ {\rm erf}\left
(\frac{\ln \ttaum - n \alpha^2 \mu_2^2 -\ln {\cal A} + \frac{\alpha \mu_2^2}{2}}
{\alpha \mu_2 \sqrt{2}} \right) \right] .
\label{taumomfin}
\end{eqnarray}
Here the relation $\mu_1 =-\mu_2^2/2$, that follows from the condition
$<\delta>=0$ for the log-normal distribution, is used.  The moments of the
truncated optical depth distribution depend on four parameters ${\cal
A}$, $\mu_2$, $\alpha$ and $\ttaum$.  The parameter $\ttaum$ is chosen
such that for $\ttau <\ttaum$ the local optical depth does not suffer
from the biases introduced in saturated regions. As apparent from
equation (\ref{taumomfin}) there are two basic degeneracies leaving
two independent parameters,
\begin{equation}
{\cal B}\equiv \ln {\cal A} - \alpha \mu_2^2/2 , \qquad \qquad 
{\cal C}\equiv\alpha \mu_2.
\end{equation}
NH00 showed that  the moments of $\ttaut$ can 
then be written in terms of these parameters as
\begin{equation}
<\ttau^n_{\rm t}>=
\frac{1}{2}\; \exp\left(\frac{n^2 {\cal C}^2} {2}+ n {\cal B} \right)
\;
\left[1+ {\rm erf}\left
(\frac{\ln \ttaum - n {\cal C}^2 - {\cal B}}
{ \sqrt{2} {\cal C}} \right) \right].
\label{taumomfin0}
\end{equation}
The first two moments, $<\ttau_{\rm t}>$ and  $<\ttau^2_{\rm t}>$,
are sufficient to determine the parameters $\cal B$ and
$\cal C$. From these one can then infer the $rms$ fluctuation 
amplitude of the gas density $\sigma_{J}$  
and the normalization constant  of the optical depth {\cA}. 
The $rms$ fluctuation  amplitude of the gas density $\sigma_{J}$ 
is related to the amplitude of the 1D power spectrum of the gas 
by a simple integration,  

\begin{equation}
\sigma^2_{J}   = \frac{1}{(2\pi)^3}\int_0^\infty P^{3D} d^3k
 =  \frac{1}{\pi} \int_0^\infty P^{1D} dk \;\; ,
\label{sig_J}
\end{equation}
where equation~\ref{eq:ps-diff} is used to obtain the equality
on the right hand side.

The normalization adds another uncertainty to the power spectrum
calculation. We estimate this uncertainty  to be $\approx 40\%$ (see NH00) for a
spectrum as long as Q1422+231. To account for the larger total 
redshift path  in each redshift bin compared to that of Q1422+231 
we assume that the amplitude errors follow a Poisson distribution,
\ie\ that  they  scale inversely with the square root of the combined 
length of the spectra in each redshift bin. This scaling
reduces the error associated with the normalization to $\sim$
10\%--20\% for each redshift bin. This error is added to the error due
to the shape measurement. Note however, that unlike the shape
measurement error, the error due to the normalization is added to all
data points equally and is highly correlated.

We have estimated the influence of the truncation at $k=10 \Mpc^{-1}h$  
on the calculation of $\sigma_J$ from equation~\ref{sig_J}
extrapolating the dependence of the power spectrum on $k$ at $k \gsim 2
\Mpc^{-1}h$. At these wavenumbers the 1D power spectrum 
scales  roughly as $k^{-3}$ (see figure~\ref{fig:taucut}). This yields a
contribution of the order of a few percent compared to the
contribution of the sampled range. Note that, since the slope of the
power spectrum is expected to further steepen at larger wavenumbers  due to
the effective Jeans scale cutoff, this is a conservative upper
limit. This will be further demonstrated with simulated spectra (see
right panel of figure~\ref{fig:1Dmeas}).

Figure~\ref{fig:PS1D} shows the measured 1D power spectra of the gas
distribution (solid curves), assuming the $\alpha$ values indicated on
each panel in the four redshift bins. The error bars shown are those
associated with the uncertainties in the shape and amplitude
measurements. The figure also shows as the dashed curves the best
fitting models obtained from the likelihood analysis, which will be
described in the following two sections.

\section{Modelling the 1D gas power spectrum}

\subsection{General Considerations} 

Equation~\ref{eq:ps-diff} can be used in order to obtain the 3D power
spectrum from the measured 1D power spectrum.  Unfortunately however,
the measured 1D power spectrum is already a noisy quantity and
differentiation will introduce more uncertainties. This approach is
explored in \S~\ref{sec:3DPS} where we  compare the
3D power spectrum obtained here to the 3D matter power
spectrum inferred from the flux power spectrum by Viel et
al. (2004). In order to quantitatively constrain the parameters
describing the matter power spectrum and the thermal state of the gas,
we use instead a likelihood analysis. 
The likelihood analysis  is described in section \ref{sec:likelihood}.  

To obtain a realistic model for the 1D power spectrum of the gas
density we start with the linear 3D matter power spectrum in real
space. From this the non-linear 3D power spectrum of the gas
distribution is obtained by taking into account the effects of
non-linearity, redshift distortions and gas \textit{vs.} dark-matter
bias.  The model for the 1D power spectrum of the gas in redshift
space is then readily obtained by using the
integral form of equation~\ref{eq:ps-diff},
\begin{equation}
P^{1\mathrm{D}}(k) = \frac{1}{2 \pi}\int_k^{\infty}
P^{3\mathrm{D}}(k') k'\mathrm{d}k'.
\label{eq:ps-int}
\end{equation}

\subsection{The Linear 3D Matter Power Spectrum} 

We restrict our  analysis to the generalized family of CDM
cosmological models, allowing variations in the mass density and
vacuum energy density parameters.  The general form of the power
spectrum of these models is,

\begin{equation}
P^{3D}(k) = A \,T^2(\Omega,\Omega_{\rm m}, \Omega_{\rm b},h; k)\, k^{n_s}. 
\label{}
\end{equation} 
The CDM transfer function, proposed by Sugiyama (1995), is adopted,
\begin{eqnarray}
T(k) & = & {\ln\left( 1+2.3q \right) \over 2.34q}\times   \nonumber\\
& & \left[1+3.89q+(16.1q)^2+(5.46q)^3+(6.71q)^4\right]^{-1/4} ~,
\label{}
\end{eqnarray}
\begin{equation}
q=k \left[ \Omega h\,
           \exp (-\Omega_{\rm b} -h^{1/2} \Omega_{\rm b}/\Omega)\,
           (h\, {\rm Mpc}^{-1} )  \right]^{-1} ~,
\label{}
\end{equation}
where $\Omega$, $\Omega_{\rm m}$ and $\Omega_{\rm b}$ are the total, mass and
baryonic density parameters,  respectively; $h$ is the Hubble
constant in units of $100 \mathrm{\kms \Mpc^{-1}}$ and $n_s$ is the 
power law index of the primordial power spectrum. The parameter $A$ is 
the power spectrum normalization factor, which will be expressed here 
in terms of the other cosmological parameters and $\sigma_8$ -- the
rms density fluctuations within top-hat spheres of $8
\mathrm{h}^{-1}\Mpc$ radius.
 
The parameters are varied such that they span a range of plausible CDM
models.  In all cases, the baryonic density is assumed to be
$\Omega_{\rm b}=0.02 h^{-2}$, the value currently favoured by
primordial nucleosynthesis analysis (\eg Walker \etal 1991, Burles \&
Tytler 1998) and the WMAP cosmic microwave background (CMB) data
(Bennet \etal 2003). The value of $h$ is fixed to $0.7$
and the Universe is assumed to be flat. 

We have focused in our study  on a few cosmological parameters,
$\Omega_{\rm m}$, $\sigma_8$ and $n_s$.  It is  important to
note that some of the free parameters of the power spectrum models are
degenerate.  For example at the range of wavenumber constrained 
by \lya forest data the amplitude of the power spectrum is degenerate with the
power spectrum power law index, $n_s$.  We have therefore 
fixed the value of the power law index to unity in most of
the analysis of the real data.  We will discuss  the influence of changing the value
of $n_s$  on the results for a few cases.

\subsection{Nonlinear effects}

As discussed earlier, the fluctuations in the \lya optical depth follow
those of the dark matter down to the Jeans scale, which is of the order
of 1 $\mathrm{h^{-1}}\Mpc$. At redshifts of 2-4, these scales have
already entered the quasi-linear regime. It is therefore necessary to account for the
non-linear evolution of the density field. Peacock and Dodds (1996) have
developed a simple recipe for mapping the linear 3D matter power
spectrum onto the quasi linear regime. The actual mapping used here is
the one described in Peacock's (1999) book which has slightly modified,
and more accurate parameters than the one given in
Peacock \& Dodds (1996).  The recipe is accurate to few percent for a
very large family of CDM power spectra. For more details, see Peacock
(1999) and references therein (see also Smith \etal 2003).

It is worth noting that it should not be necessary to follow the evolution of
the power spectrum to the highly non-linear regime. High density
regions will not give rise to the low column density spectral lines we
use in our analysis. The influence of saturated lines is further 
diminished by introducing a severe cutoff to the optical depth (see
figure~\ref{fig:taucut}).

\begin{figure*}
\setlength{\unitlength}{1 cm} \centering
\begin{picture}(17,12)
\put(-1.9, -1.){\includegraphics{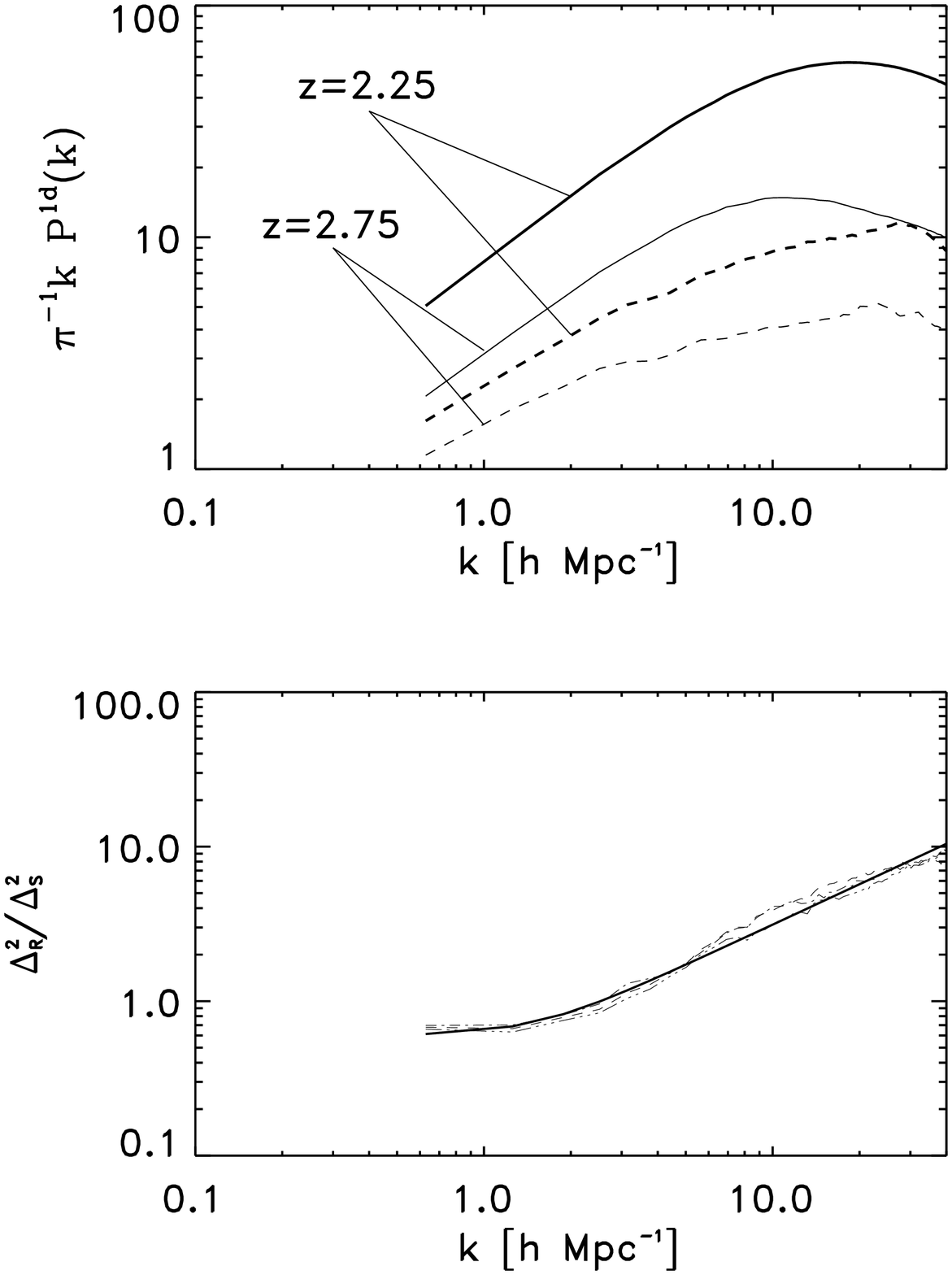}}
\put(7.9, -1.){\includegraphics{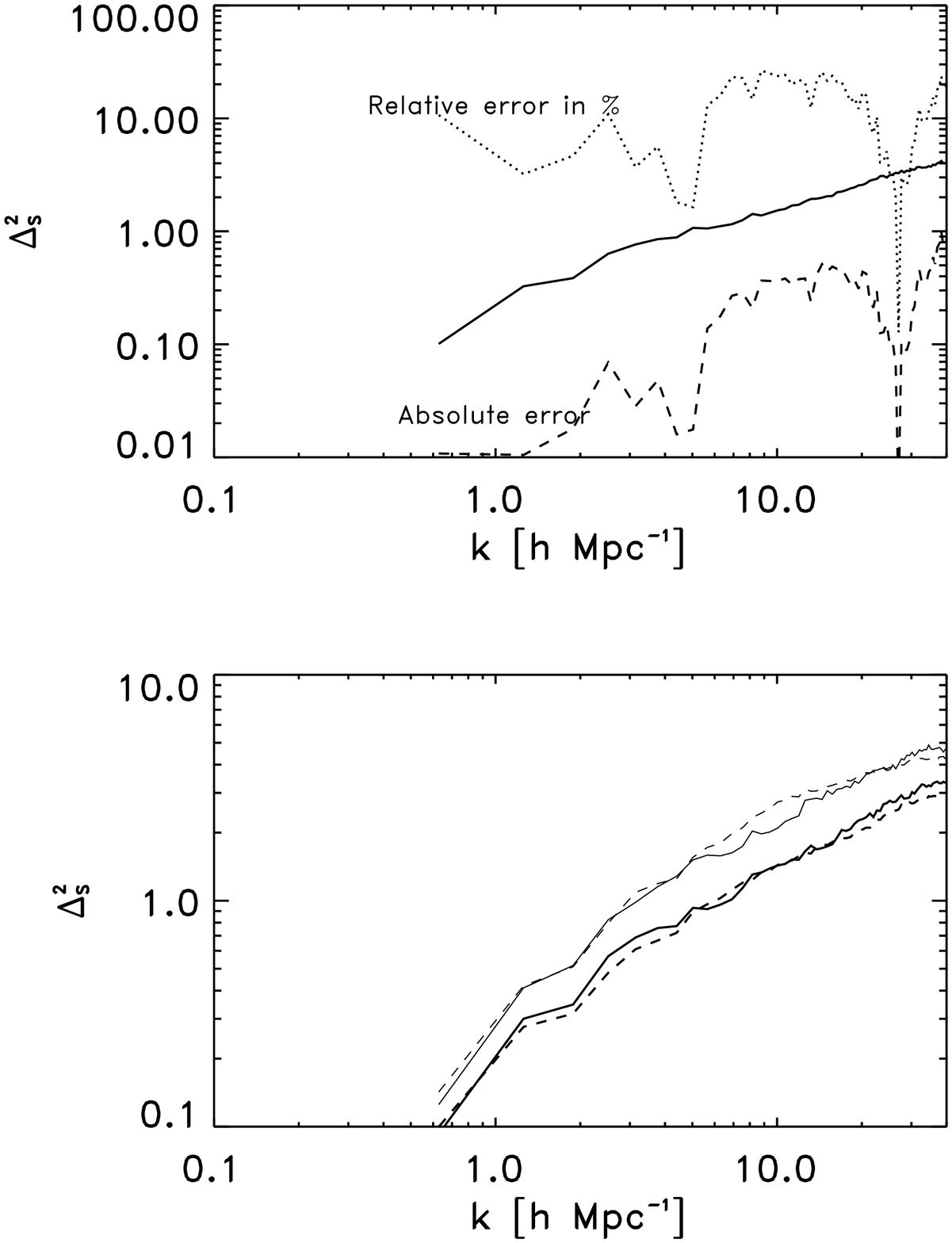}}
\end{picture}
\caption{Redshift distortions of the 1D and 3D power spectra  and
their errors as estimated from the numerical simulations. The upper
left panel shows the real-space (solid curves) and redshift-space (dashed
curves) 1D power spectra at  two redshifts, $2.75$ (thin) and
$2.25$ (thick). The lower left panel shows the ratio between the
dimensionless real space 3D power spectrum and its redshift space
counterpart for three redshifts, $2.25$, $2.75$ and $3$  as
dashed/dotted curves. The solid curves  shows the fit to the ratio
used in the likelihood analysis. The upper right panel shows the
errors introduced in the dimensionless 3D redshift-space  power
spectrum by using this fit. The solid curves shows the dimensionless 3D
power spectrum at $z=2.75$; the dashed curve shows the absolute error
and the dashed curve shows the relative error which has a maximum value
of about $20\%$. The errors at the other two redshifts are
similar. The lower right panel shows the actual 3D real space  power
spectra (solid curves) \textit{vs.} the reconstructed
power spectra (dashed curves) at  two redshifts, $2.75$ (thick) and
$2.25$ (thin).}
\label{fig:zdist}
\end{figure*}

\subsection{Redshift distortions}
\label{sec:distortions}

After taking the non-linear effects into account, the theoretical 3D
real-space power spectrum is transformed into redshift space. The ``distortions''
caused by this transformation -- normally called redshift distortions
-- will significantly bias the power spectrum measurement. On linear
scales, redshift distortions cause an enhancement in the measured
power spectrum with a constant factor that depends on the value of
$\Omega_{\rm m}$ at a given redshift (Kaiser 1987).  In the highly
non-linear regime, redshift distortions tend to dilute the distribution
along the line of sight and create the so called ``fingers of God''.
Redshift distortion in the highly non-linear regime thus lead to 
a suppression of the power spectrum relative to the
real-space power spectrum. The effect of peculiar velocities in
this regime can be modelled in a statistical manner with a Gaussian fit to the 1D
pairwise velocity distribution (Davis \& Peebles 1983). In the case at
hand however, most of density fluctuations are in the linear to
quasi-linear regimes. Unfortunately, there is no good analytical
description of  redshift distortions in the quasi-linear regime. 

To account for the redshift distortions we have thus used the
hydrodynamical simulations described in section
\S~\ref{sec:numerical} to determine an empirical relation between
the real-space and redshift power spectra (see
figure~\ref{fig:zdist}). Note that this correction is dynamical and as
such is not dependent on the details of the gas physics.

The upper left panel of figure~\ref{fig:zdist} shows the 
dimensionless 1D power spectra with (dashed curves) and without (solid
curves) peculiar velocity distortions. The magnitude of the effect may
appear rather large, however one should keep in mind that the 1D power
spectrum is an integral quantity (Eq.~\ref{eq:ps-int}) which 
tends to enhance any systematic biases in the 3D power spectrum.

To avoid this accumulation of bias in the 1D power spectrum, the
influence of the velocity distortions is modelled using the 3D power
spectrum. The lower left panel of figure~\ref{fig:zdist} shows the
ratio between the real-space and redshift space 3D power spectra in
the relevant range of wavenumbers as deduced from the numerical
simulations. The ratio is shown for 3 different redshifts.  At the
largest scales the ratio flattens to a constant value that is
comparable to  the linear effect described by Kaiser (1987). On smaller
scales the trend reverses and the redshift space power spectrum is
suppressed relative to the real-space power spectrum. We fit the ratio between the two
with the following simple formula,
\begin{equation}
\frac{P_r(k)}{P_s(k)} = 
 0.535\, \left({1 +(k/k_0)^3}\right)^\frac{1}{3},
\label{eq:zdist}
\end{equation}
where $P_r(k)$ and $P_s(k)$ are the real-space and redshift space
power spectra respectively;  $k_0=2 \mathrm{Mpc^{-1}h}$ is the 
scale relevant for the transition from linear to quasi-linear regime. 
The fit is shown with the  solid curve. From figure~\ref{fig:zdist} it is clear 
that the required correction for the ratio is almost independent of
redshift. A correction with the same functional fit can thus be applied 
to all redshifts. The errors introduced by the fitting formula are
shown in the upper right panel. The relative error amounts to a
maximum of about 20\%. The lower right panel shows how well the
correction recovers the real 3D redshift space power spectrum.

For flat cosmological models, the Universe at the redshifts of the 
\lya spectra is very similar to an Einstein-de-Sitter
model. The correction proposed in equation~\ref{eq:zdist} will,
therefore, not be very sensitive to the exact values of cosmological
parameters, possibly with the exception of $\sigma_8$. We have shown
that  the relation holds at three different redshifts. The functional form of
equation~\ref{eq:zdist} should thus be a good fit which depends only 
weakly on cosmological parameters.

\subsection{Gas vs. Dark-Matter Power Spectra}

The next issue is to model the bias between the gas and dark matter
distribution.  On scales larger than the Jeans scale, numerical
simulations have shown that pressure effects are, as expected,
negligible. The gas faithfully traces the
dark matter on these scales. On scales comparable or smaller than the Jeans scale,
however, the distribution of gas deviates from that of the
dark matter. At these scales, the pressure becomes important and prevents
the gas from contracting, smoothing out all the small scale fluctuations.

\begin{figure*}
\setlength{\unitlength}{1cm} \centering
\begin{picture}(16,11)
\put(2.5, -1){\includegraphics{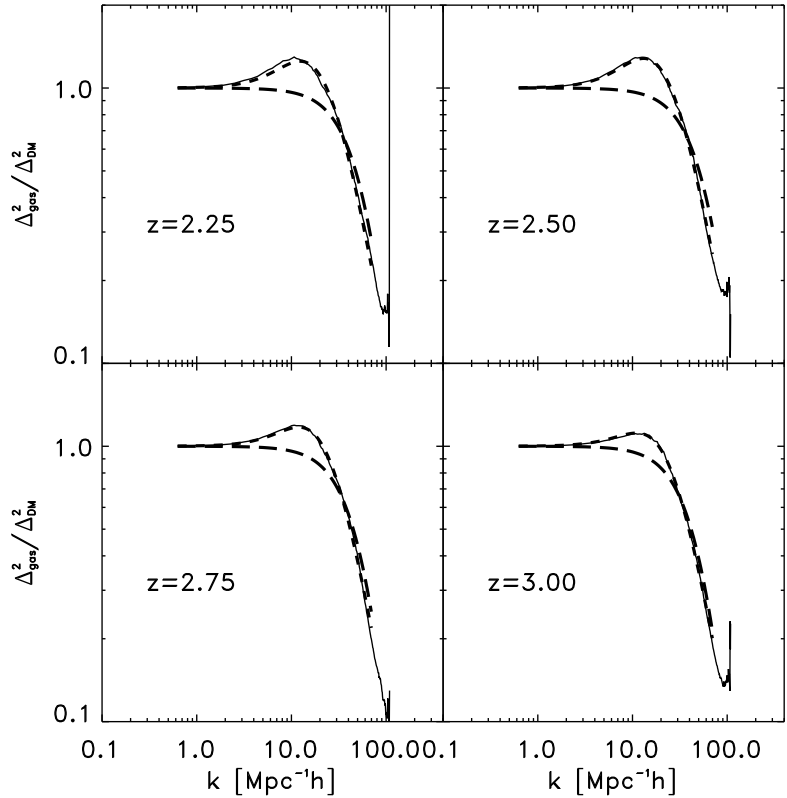}}
\end{picture}
\caption{ Gas \textit{vs.} dark matter power spectra. The bias between
the gas and the dark matter due to Jeans smoothing is shown. The solid
curve shows the ratio of the gas and dark-matter 3D real space
power spectra. The short dashed curve is the fitting function used in
the current analysis. The long-dashed curve is the normal Jeans scale
cutoff function. Note that the hydro-simulations used here do not
include gas cooling (see text). }
\label{fig:gasDM}
\end{figure*}

An approximate functional form for the effect of Jeans smoothing on
the power spectrum can be calculated from linear theory ({\it e.g.}
NH99).  This is, however,  not adequate to describe the
quasi-linear regime, relevant for this study. We have therefore again
used the  numerical simulations described in
\S~\ref{sec:numerical} to obtain a more accurate fit to the ratio
between the power spectra of gas and
dark matter densities.  The solid curves in figure~\ref{fig:gasDM}
show this ratio for four redshift bins as measured directly from the
simulations.  Notice the slight enhancement at $k\approx 20
\mathrm{Mpc^{-1}h}$.  The long-dashed curves are the best fit
functions based on linear theory, which do not
reproduce the enhancement.  The dashed curves show the results of fitting
the simulation data with a fitting function of the form,
\begin{equation}
P^{3D}_{gas} = \frac{P^{3D}_{DM}\left(1 + B(z)\, k^2\right)} {\left(
1+\left( \frac{k X_J}{2 \pi}\right)^2\right)^{2}},
\end{equation}
where $P^{3D}_{gas}$ and $P^{3D}_{DM}$ are the 3D gas and dark-matter
power spectra respectively, and $X_J$ is the effective Jeans scale,
and $B(z) = 0.1447 - 0.0186 \, z$. The function $B(z)$ is later used in
the likelihood analysis where the value of $X_J$ was kept as a free
parameter.

It worth noting that we have used simulations with and without
cooling.  The main difference we found is that simulations without
cooling require the addition of the term, $B(z)\, k^2$, to the
fit. Whereas the simulations with cooling require no such addition to
the normal Jeans cutoff function. In addition to that, we suspect that
the SPH simulations with cooling-functions suffer from excess cooling
of the gas, a well known issue in SPH simulations. For those reasons,
we adopt the more general fit that the hydrodynamical simulations
without gas cooling produce.

\subsection{Testing the recovery of the 1D power spectrum 
from the optical depth with numerical simulations} 
\label{sec:numerical}

In order to test how well our likelihood analysis recovers the
non-linear 1D gas power spectrum in redshift space we have tested it
with artificial spectra produced from state-of-the art SPH simulations.  
A suite of simulations with varying particle numbers, resolution and
boxsize have been  carried out with the
parallel TreeSPH code {\small GADGET-2} (Springel, Yoshida \& White,
2001; Springel 2005).  {\small GADGET-2} was used in its TreePM mode
which speeds up the calculation of long-range gravitational forces
considerably. The simulations were performed with periodic boundary
conditions with an equal number of dark matter and gas particles and
used the conservative `entropy-formulation' of SPH proposed by
Springel \& Hernquist (2002).
The mean UV background produced by quasars 
as calculated by  Haardt \& Madau (1996) has been assumed.  
This leads to reionization of the Universe at $z\simeq
6$. The simulations were run with the equilibrium solver for the
thermal and ionization state implemented in {\small GADGET-2}. The
heating rates at $z>3.2$ were increased by a factor of 3.3 in order to
achieve temperatures which are close to observed temperatures (Schaye
et al. 2000, Theuns et al. 2002, Ricotti, Gnedin \& Shull 2000).  At
$z<6$ the power-law index of the gas density temperature relation is
$\gamma \sim 1.6$, where $T=T_0\,(1+\delta)^{\gamma(z)-1}$.

To maximise the speed of the simulations, a simplified star-formation
criterion in the majority of the runs is employed. All gas at densities larger
than 1000 times the mean density was turned into collisionless
stars. The absorption systems producing the \lya forest have small
overdensity so this criterion has little effect on flux statistics,
while speeding up the calculation by a factor of $\sim 6$.
All feedback options of {\small GADGET-2} in the simulations have been
turned off.

We have run four simulations with a box size of 60, 30, 15 and 10
comoving $h^{-1}\,$ Mpc, respectively. The three larger simulations
were run with $2\times 400^3$ particles including gas cooling.  The
simulation with a box size of 10 comoving $h^{-1}$ Mpc was run with
$2\times 200^3$ particles and without radiative cooling. This
was done in the smallest box size simulations only in order to address 
the effect of the Jeans smoothing without allowing the gas to
radiatively cool. In these simulations the thermal state of the gas is 
set by the equilibrium between photo-heating and adiabatic cooling caused by the
expansion of the Universe. The cosmological parameters were chosen to 
be consistent with the values obtained by the WMAP team in their
analysis of WMAP and other data
(Spergel et al. 2003), $\Omega_{{\rm m}}= 0.26$, $\Omega_{\Lambda} =
0.74$, $\Omega_{{\rm b}} = 0.0463$ and $H_0=72\,{\rm
km\,s^{-1}Mpc^{-1}}$.  The CDM transfer functions of all models have
been taken from Eisenstein \& Hu (1999).

\begin{figure*}
\setlength{\unitlength}{1cm} \centering
\begin{picture}(8,12.5)
\put(-1.3,-1.5){\includegraphics{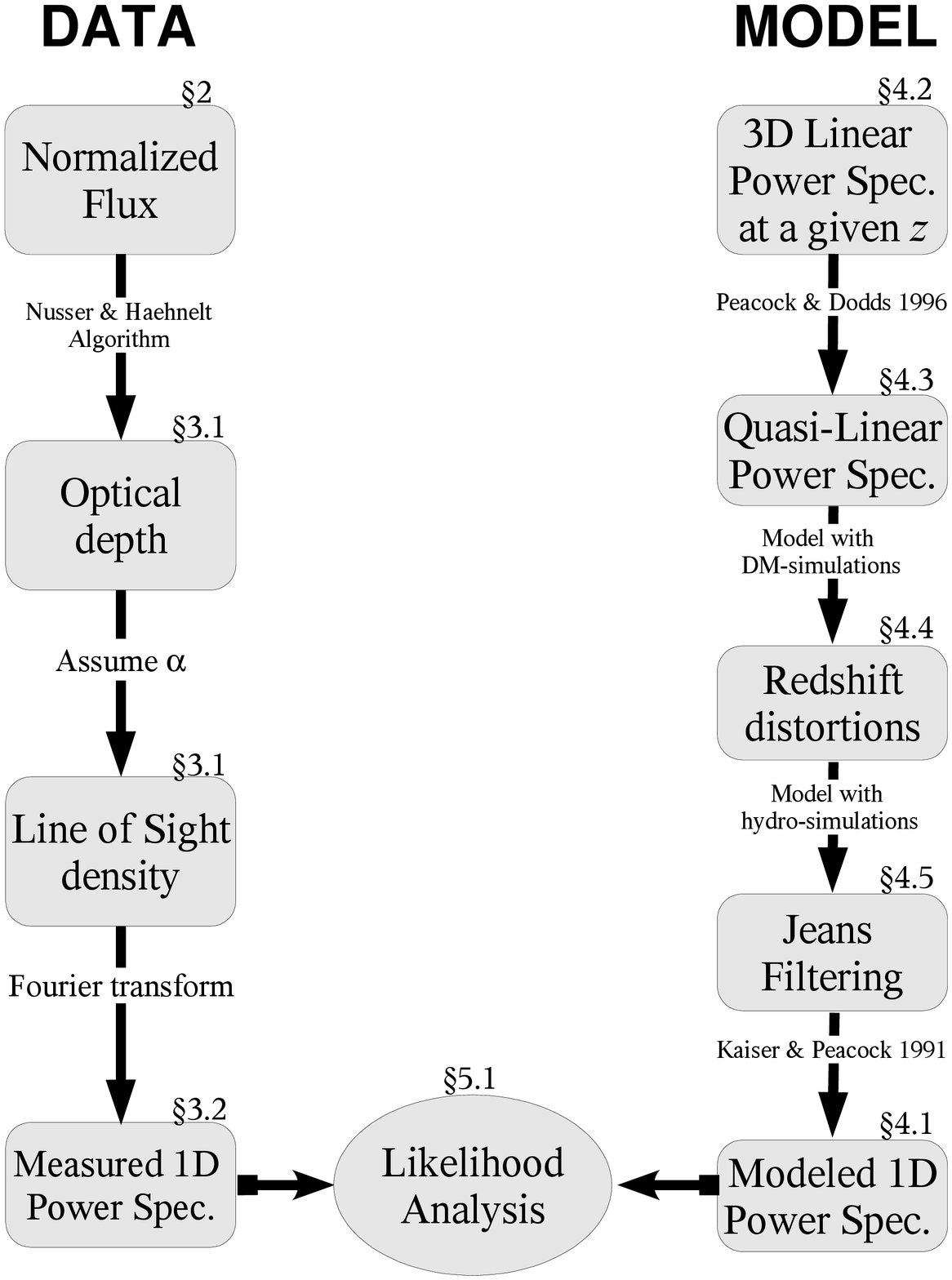}}
\end{picture}
\caption{A flow-chart summarising the analysis steps as applied to the
data and the model. The sections in the paper in which each step is
explained are indicated at the upper right corner of each ``step-box''.}
\label{fig:chart}
\end{figure*}


\begin{figure*}
\setlength{\unitlength}{1cm} \centering
\begin{picture}(17,8.5)
\put(-1., -1.5){\includegraphics{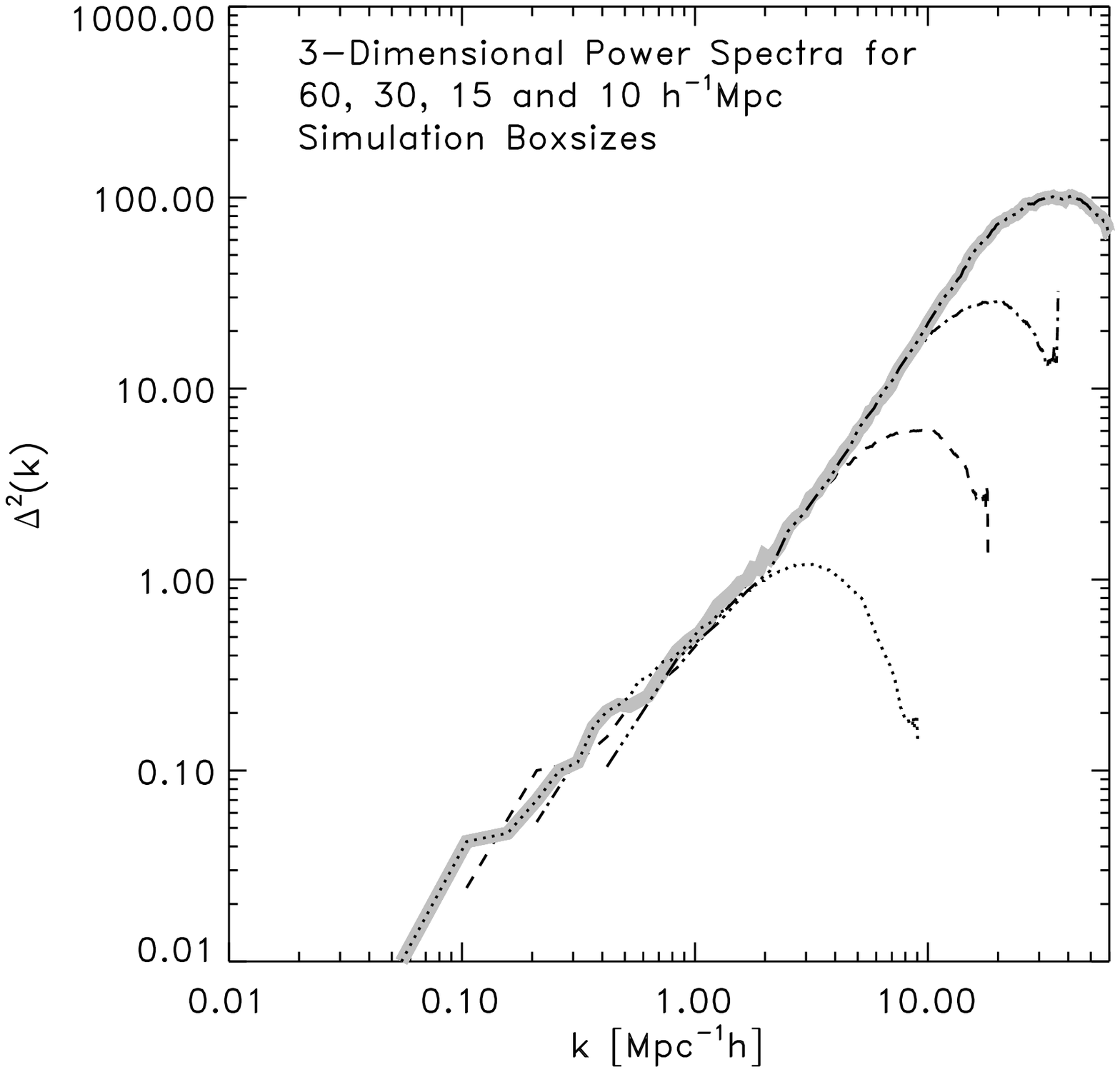}}
\put(8.2, -1.5){\includegraphics{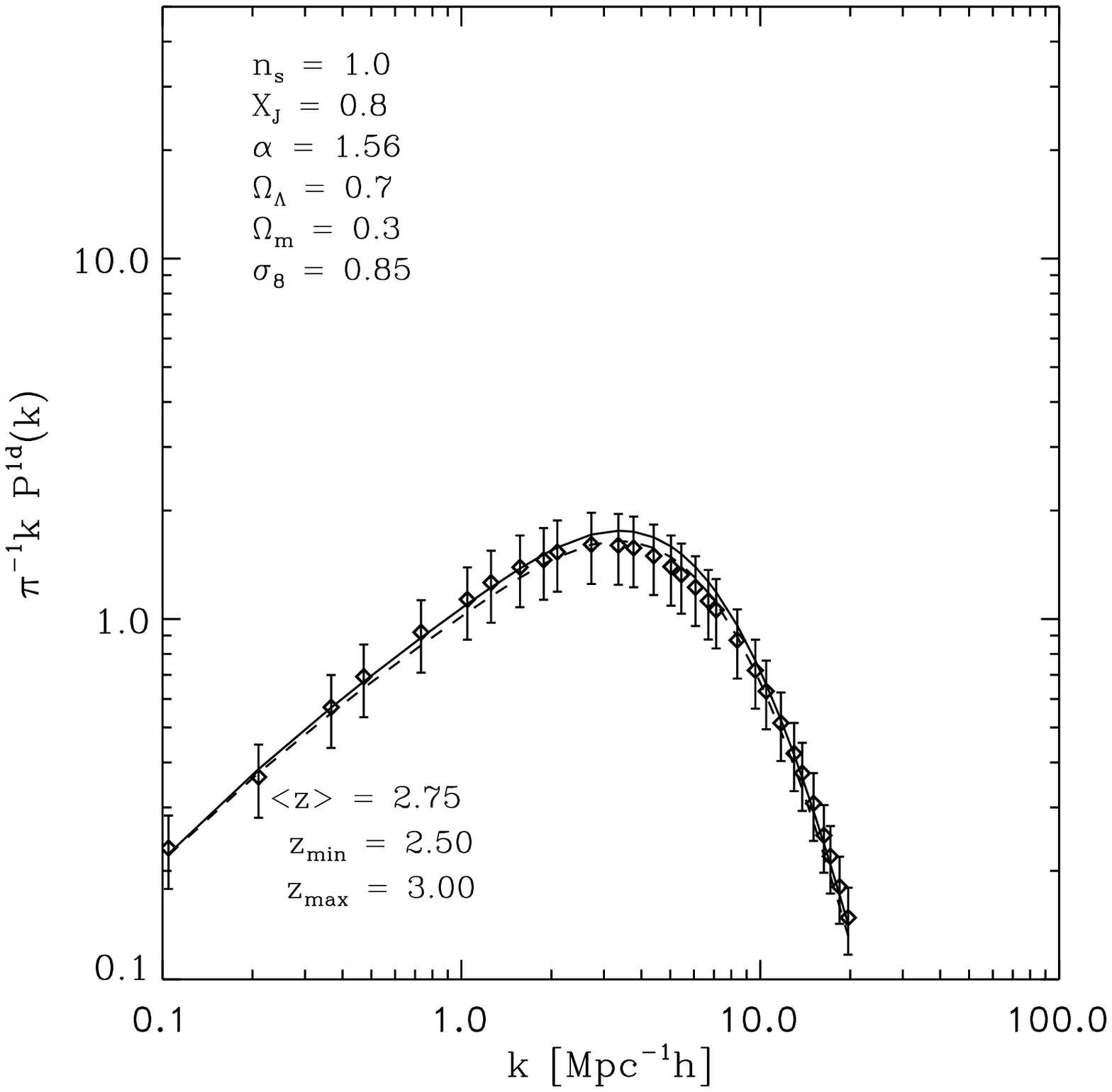}}
\end{picture}
\caption{{\it Left panel:} shows the 3D power spectra of the gas density
as measured from the simulations with a box size of 60 (dotted), 30
(dashed), 15 (dotted-dashed), and 10 (triple-dotted-dashed) $h^{-1}$
Mpc boxsizes. Clearly, none of the simulations spans the dynamical
range needed to capture simultaneously the large and small scales
needed to model the 1D power spectrum. The grey solid curve is the
composite 3D power spectrum which is our best guess of what the
non-linear 3D power should look like with infinite dynamic
range. {\it Right panel:} the diamond symbols show the 1D power spectrum of
the gas density corresponding to the composite 3D power spectrum in
the left panel (with error bars) with a Jeans cutoff at $X_J=0.8
\mathrm{h^{-1} Mpc}$. The solid curves shows the best fit power
spectrum found by the likelihood analysis with cosmological parameters
as annotated on the figure. The dashed curve shows the 1D
power spectrum with the actual parameters of the hydro-simulations with
$X_J=0.8 \mathrm{h^{-1} Mpc}$.}
\label{fig:1Dmeas}
\end{figure*}

The left panel of Fig.~\ref{fig:1Dmeas} shows the 3D gas power
spectrum for the simulation with a box size of 60 (dotted), 30
(dashed), 15 (dotted-dashed), and 10 (triple-dotted-dashed) $h^{-1}\,$
Mpc.  Note that the hydro-simulation have insufficient dynamic range
to simulate the forest down to the Jeans scale and capture at the same
time the largest  structure probed by  the observed \lya forest. 
Unfortunately, such a dynamical range is currently
inaccessible with a single simulation. We have therefore combined 
the 3D power spectra of the four simulations  to obtain an estimate of a power
spectrum that spans scales that are comparable to those probed by the
data. Note that this approach can not be applied directly to the 1D
power spectra of the simulations due to the integral nature of the 1D
power spectrum. The best estimate of the 1D power spectrum expected
for infinite dynamic range is, thus, obtained with
equation~\ref{eq:ps-int} using the combined 3D power spectrum of the
gas obtained from the simulations. This mock 1D power spectrum is
calculated from the simulation outputs at $z=2.75$. We have added
errors that are comparable to those found in the data.

The right panel of Fig.~\ref{fig:1Dmeas} shows the dimensionless 1D
power spectrum obtained by applying an artificial Jeans-like cutoff 
with $X_J=0.8 \mathrm{h^{-1} Mpc}$  to the 1D power spectrum obtained
from the combined 3D power spectrum from the simulations
(diamond symbols with error bars).  The dashed
curve shows the non-linear analytical model for the same cosmological
parameters and amplitude of the linear matter power spectrum as
implemented in the hydro-simulation with a Jeans cutoff at $X_J=0.8
\mathrm{h^{-1} Mpc}$. The agreement between the non-linear 1D power
spectrum model of the gas density and our best estimate from the set
of hydro-simulations is, perhaps not too surprisingly, excellent.

 \subsection{Summary of the analysis steps}

To summarize the analysis steps: First, the non-linear 1D power
spectrum of the gas distribution is modelled analytically. In this way
we can easily vary the shape and amplitude of the underlying linear
matter power spectrum. The thermal state of the gas is modelled by a
thermal smoothing function which has one free parameter, the effective
Jeans length. The shape of this function is derived from our
hydrodynamical simulations. Second, the model 1D power spectra are
compared with those recovered from the truncated optical depth of the
\lya forest data using a maximum likelihood analysis.  These steps are
summarized in the flow-chart shown in Fig.~\ref{fig:chart}

\section{Constraining the parameters of the matter power spectrum and 
         the thermal state of the gas}

\label{sec:parameters}

\subsection{The likelihood method}  
\label{sec:likelihood}

For this the following likelihood is maximized,
\begin{equation}
{\cal{L}} = const. \times e^{-\frac{1}{2}\left(\mathbf{P}^{1D}_{\rm obs} -
\mathbf{P}^{1D}_{\rm model}\right)^+ \mathbf{C}^{-1}
\left(\mathbf{P}^{1D}_{\rm obs} -
\mathbf{P}^{1D}_{\rm model}\right)}  \;\; ,
\end{equation}
where $\mathbf{P}^{1D}_{\rm obs}$ and $\mathbf{P}^{1D}_{\rm model}$ are the
observed and model power spectrum vectors and $\mathbf{C}$ is the
error correlation function.

As mentioned earlier, given the large uncertainties in the data and
the various degeneracies in the models (\eg\ between the power law
index, $n_s$ and $\sigma_8$) we will not attempt to simultaneously
constrain all free parameters on which the power spectrum models
depend.  Instead, the analysis will focus on three or less free
parameters at a time while keeping the other parameters fixed.

\subsection{Error estimates}
\label{sec:error}

The errors used in the likelihood analysis come from two sources, the
1D power spectrum measurement and the uncertainties in the theoretical
model used to fit it. The error in the measurement were discussed
earlier in \S~\ref{obs_1D_PS}. These were shown to have an
uncorrelated contribution due to the power spectrum shape
determination and a correlated component due to the amplitude
determination.

The main source of error in the model comes from the fit used to
account for the redshift distortions which introduces a error that can
be as high as $20\%$. To simplify the treatment we
conservatively  fix the relative error to be  $20\%$. This error is
added coherently to all wavenumbers.

To summarize, the error matrix, $\mathbf{C}$, is defined as follows:
\begin{equation}
\mathbf{C} = \mathit{diag}(\Delta^2_{\rm shape}) +
\left(\epsilon^2_{\rm amplitude}+ \epsilon_{\mathrm{z}-{\rm distortions}}^2
\right)\mathbf{P}_{\rm obs} \mathbf{P^+}_{\rm obs},
\end{equation}
where $\Delta_{\rm shape}$ is the error introduced by the power
spectrum shape measurement. We performed a covariance analysis which showed that, for a given 
choice of wavenumbers $k$, this error is not correlated across different
wavenumbers. $\epsilon_{\rm amplitude}$ is the relative error due to
the amplitude measurement and $\epsilon_{\mathrm{z}-{\rm
distortions}}$ is the error produced by the correction for redshift
distortions. Both errors are highly correlated.

Typical values of $\Delta_{\rm shape}$ are shown in
figure~\ref{fig:taucut} and amount on average to about $50\%$ at each
$k$ (see section \S~\ref{obs_1D_PS} for a detailed description).  The
value of $\epsilon_{\mathrm{z}-{\rm distortions}}$ is taken to be
$20\%$.  The value of $\epsilon_{\rm amplitude}$ is taken
to be $20\%$ for a spectrum of the same length as that of Q1422+231
and is assumed to scale with the inverse of the square-root
of the total length of the spectrum used in each redshift bin. This 
scaling of the errors assumes a Poisson distribution of the \lya absorption
features.

\subsection{Testing the likelihood analysis with the hydro-simulations}
\label{sec:test}
\begin{figure*}
\setlength{\unitlength}{1cm} \centering
\begin{picture}(16,13)
\put(18.,0){\includegraphics{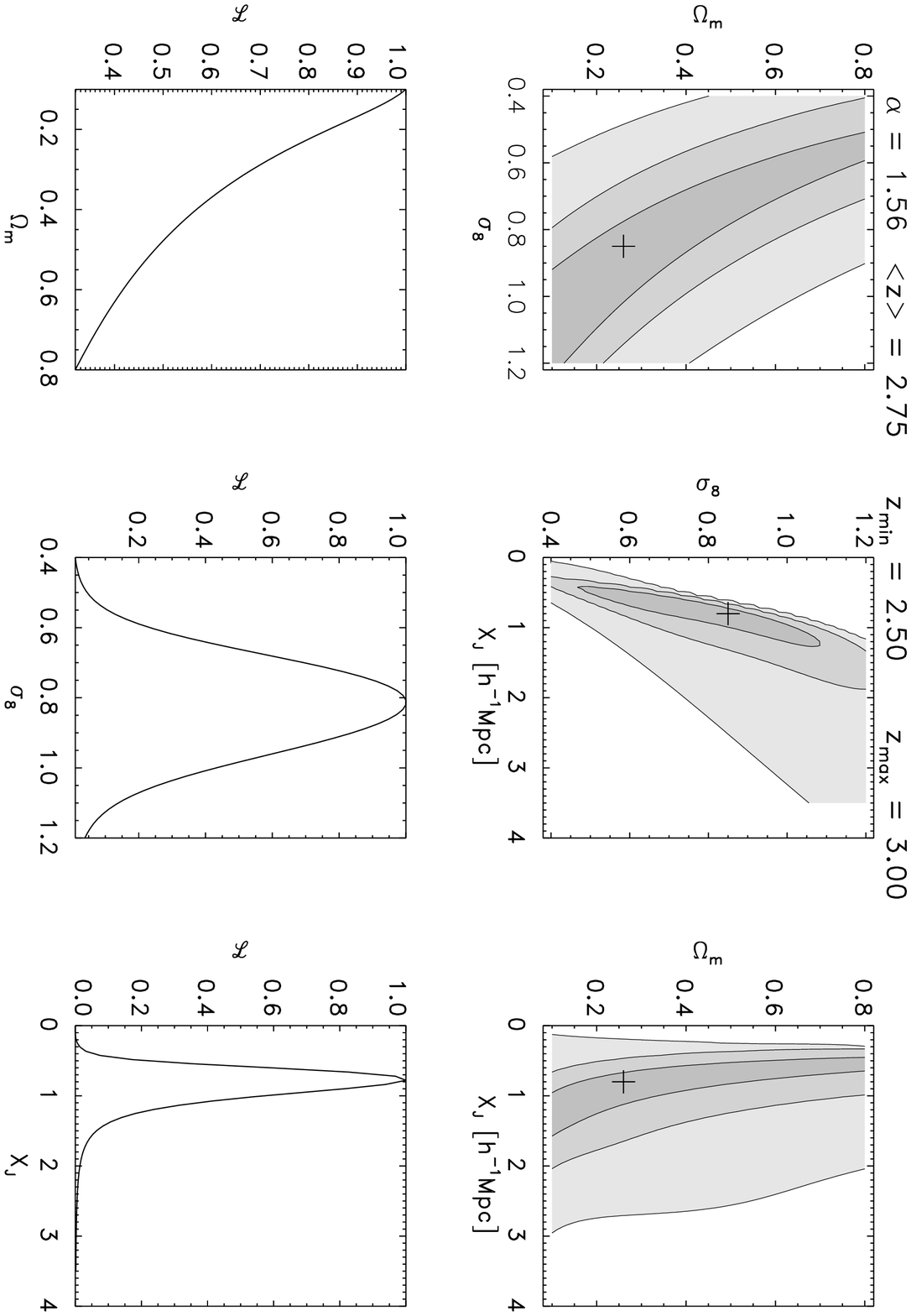}}
\end{picture}
\caption{The likelihood results for the power spectrum from the
hydrodynamical simulations with $\Omega_{\rm m}$, $\sigma_8$ and $X_J$ 
as free parameters. The upper three figure show the 1, 2 and 3-$\sigma$ 
likelihood contours for each pair of the parameters after
marginalizing over the third. The lower three panels show the 
likelihood function for each of the three parameters (after
marginalizing over the other two). The likelihood is normalized 
to have a maximum value of unity. The value of $\Omega_{\rm m}$ is poorly 
constrained, but the values of $\sigma_8$ and and $X_J$ are well 
constrained and agree with the simulations parameters.
}
\label{fig:like_sim}
\end{figure*}

In order to test how well our  procedure of constraining the model
parameter works, we have performed a likelihood analysis for a sample
of  artificial spectra obtained from the hydrodynamical simulations.  
The numerical simulation was for a flat $\Lambda$CDM model with
$\Omega_{\rm m}=0.26$, $\sigma_{8} = 0.85$, $n_s=1$, and has a
temperature-density relation that corresponds to $\alpha = 1.56$.  For the
likelihood analysis, the Universe is assumed to be flat and that
$\alpha=1.56$.  This leaves $\Omega_{\rm m}$, $\sigma_{8}$ and the
effective Jeans length $X_{J}$ as free parameters.

For the analysis of the hydro-simulation, the ``composite'' non-linear
1D power spectrum, shown in right panel of Fig.~\ref{fig:1Dmeas}, is
used.  Figure~\ref{fig:like_sim} shows three two dimensional
likelihood contours (with marginalization over the third parameters)
and the one dimensional likelihood (normalized to have a maximum of
unity) for each of the parameters with the other two marginalized
over. The top left panel shows $\Omega_{\rm m}$ {\it vs.}
$\sigma_{8}$.  The actual values are denoted by a cross.  For the
correct value of $\Omega_{\rm m}$ the actual value of $\sigma_{8}$
falls well into the 1-$\sigma$ limits of the recovered value
suggesting that the method works fine.  Note, however, that there is a
degeneracy of the inferred $\sigma_{8}$ with the assumed $\Omega_{\rm
m}$. Not surprisingly the \lya forest data alone cannot constrain both
parameters. The top middle and right panel of
Figure~\ref{fig:like_sim} show the likelihood contours of the
effective Jeans length $X_{J}$ {\it vs} $\sigma_{8}$ and $\Omega_{\rm
m}$, respectively.  Due to the cut-off of the power spectrum at small
scales, $X_{J}$ is well constrained.

The solid curve at the right panel of Fig.~\ref{fig:1Dmeas}, shows the
non-linear model power spectrum for the best fit parameters obtained
with the likelihood analysis (see \S\ref{sec:parameters}). The
agreement with the best estimate from the hydro-simulations is again
very good. The recipes used to model the non-linear evolution of the
matter power spectrum, the bias between dark matter and gas density
and the recovery of the parameters describing the optical depth
distribution, appear to work very well.

We have also tested how sensitive our analysis is to the functional
form which we use to model the Jeans smoothing by setting $B(z)=0$.  
We found no significant difference with regard to the deduced cosmological
parameters except on the value of $X_J$ itself (for more discussion on
this see section~\ref{sec:discuss}).  The reason is that even with the
canonical functional form for the Jean's smoothing deviations
from the actual ratio are $\approx 15\%$ within the observed $k$
range.  This is small compared to other uncertainties in the modelling.

\begin{figure*}
\setlength{\unitlength}{1cm} \centering
\begin{picture}(16,13)
\put(18.,0){\includegraphics{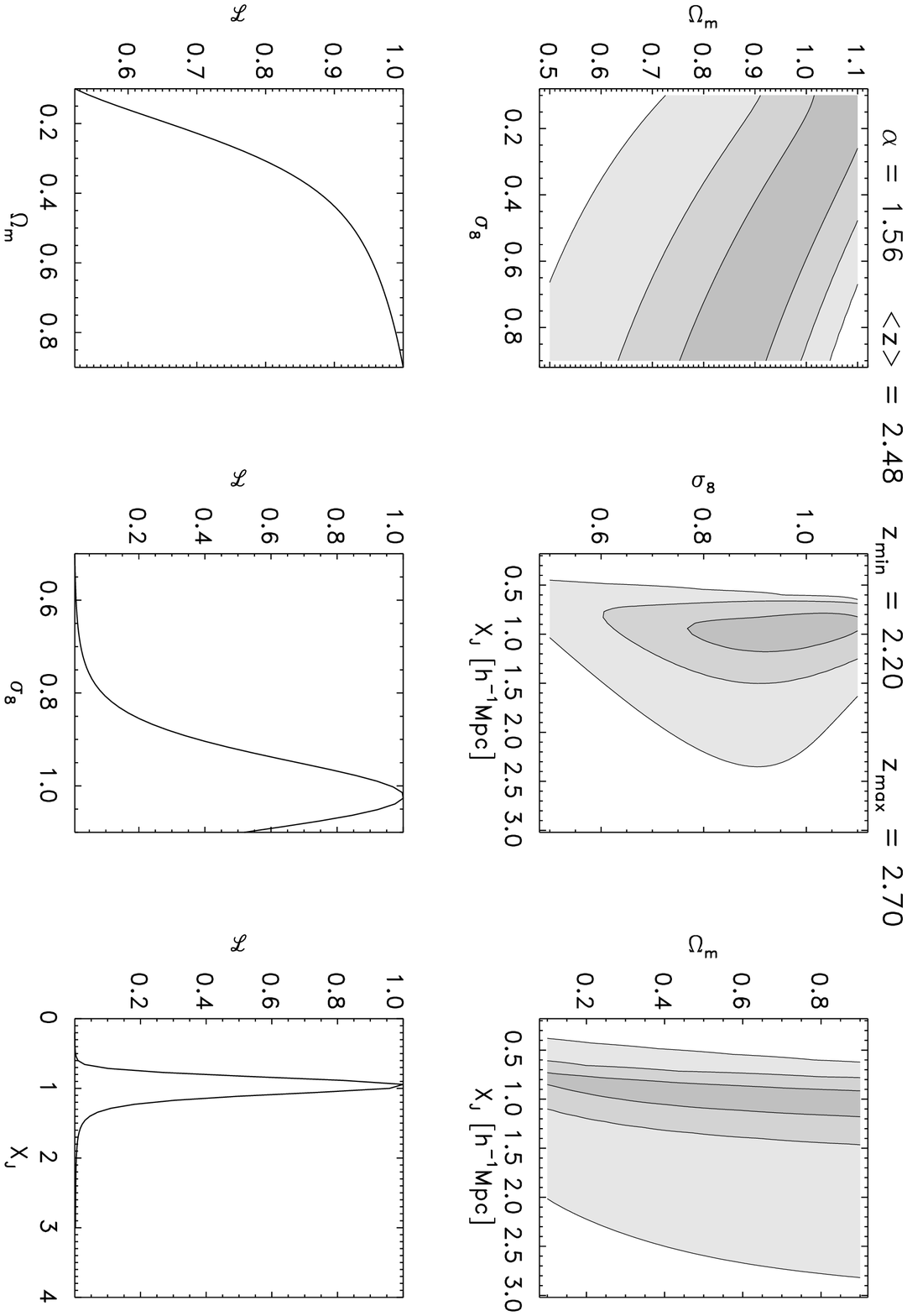}}
\end{picture}
\caption{The likelihood results for the observed spectra for the
redshift bin at $z=2.48$
with $\Omega_{\rm m}$, $\sigma_8$ and $X_J$ as free parameters.  The
assumed value of $\alpha$ is $1.56$. The upper three figures show the 1, 2 and
3-$\sigma$ likelihood contours for each pair of the parameters after
marginalizing over the third.  The value of $\Omega_{\rm m}$ is poorly
constrained, but the values of $\sigma_8$ and and $X_J$ are well
constrained.}
\label{fig:like_xos_z24}
\end{figure*}

\begin{figure*}
\setlength{\unitlength}{1cm} \centering
\begin{picture}(16,13)
\put(18.,0){\includegraphics{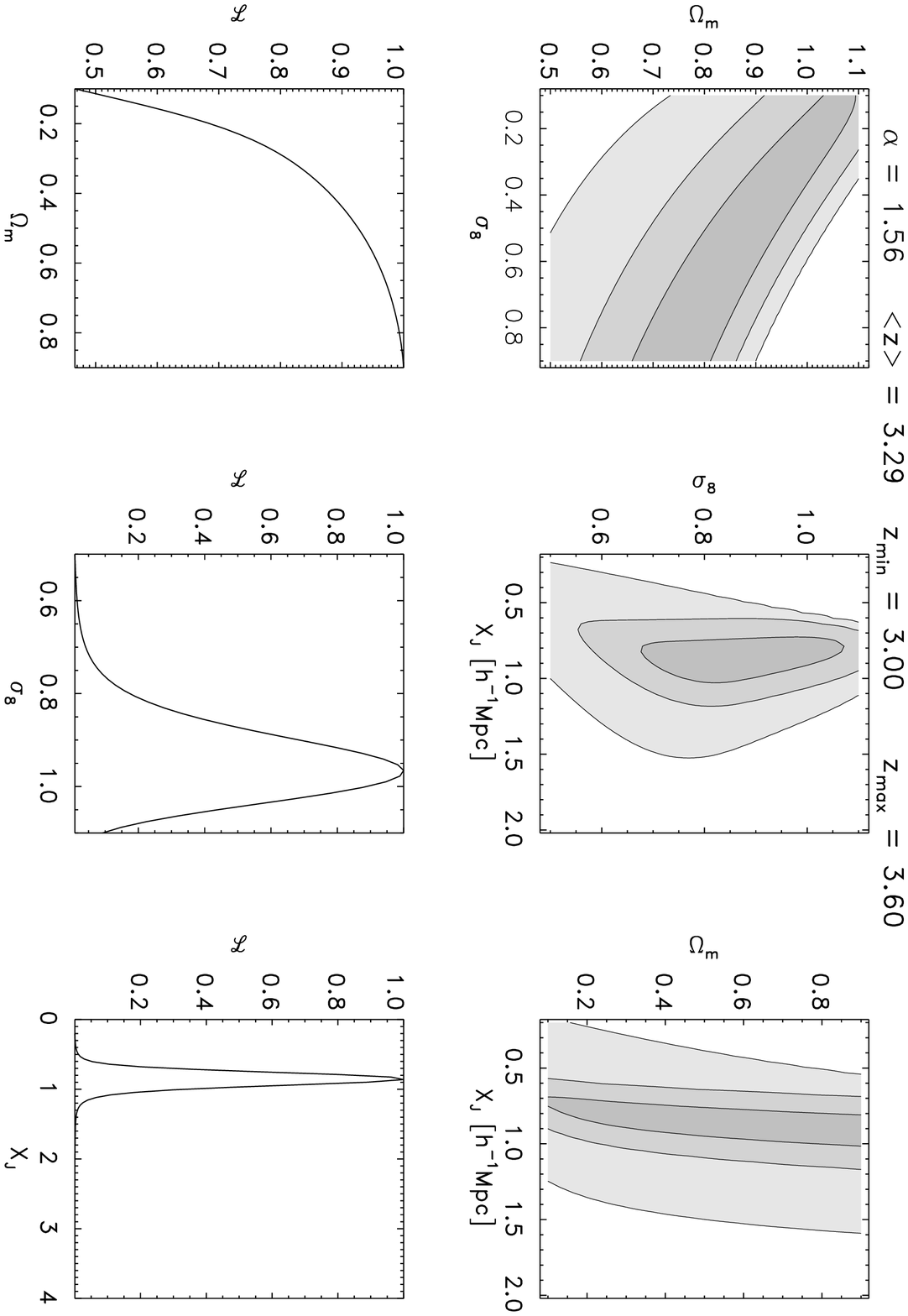}}
\end{picture}
\caption{ The same as figure~\ref{fig:like_xos_z24} but for the redshift
bin at  $z=3.29$.  The preferred value of $\sigma_8$ and $X_J$ here are
lower than those of figure~\ref{fig:like_xos_z24}
suggesting a change of $\alpha$.  }
\label{fig:like_xos_z32}
\end{figure*}

\subsection{Constraints on $\sigma_8$, $X_J$ and $\Omega_{\rm m}$ for a flat cosmology}
\label{cosmopar}
We now turn to a likelihood analysis of the real data. One difficulty is that
we will have to assume  a value of $\alpha$, the power law index used
in equation~\ref{eq:lod}, which relates the local optical depth with 
the underlying density. However as we will see later, by comparing 
results of different redshifts, we can actually constrain the evolution
of $\alpha$ and thus the thermal evolution of the IGM.  As discussed 
earlier, $\alpha$ ranges between $1.56-2$, with the lower
limit corresponding to the balance between  adiabatic cooling and
photoheating and the upper limit to an isothermal IGM. During a phase
in which the IGM is rapidly heated (\eg\ during the reionization 
the value of $\alpha$ becomes closer to 2).  

Equation~\ref{eq:lod} suggests that for a given value of the local
optical depth, $\ttau$, the amplitude of the density will be larger
for smaller $\alpha$ and vice versa.  In this subsection, $\alpha$ is
chosen to be either $1.56$ or $1.8$, for all redshift bins. Also shown
are results for a `mixed' case in which $\alpha$ evolves with redshift
as expected for a scenario in which \hep\ is ionised to \hepp\ at
$z\approx 3.2$.

As discussed above, the value of $n_s$ is degenerate with $\sigma_8$,
and we have assumed $n_s$ to be fixed to a value of
unity. We have furthermore restricted our analysis to models with a flat
cosmology (\ie\ $\Omega_{\rm m}+\Omega_\Lambda =1$).  The free
parameters of this analysis are $\Omega_{\rm m}$, $X_J$ and
$\sigma_8$.

Figures~\ref{fig:like_xos_z24}
and~\ref{fig:like_xos_z32} show the results of our analysis assuming
$\alpha=1.56$ at $z=2.48$ and $z=3.29$ respectively. The
same three two-dimensional likelihood contours and the one dimensional
likelihood curves as in the test with numerical simulations are shown
in fig.~\ref{fig:like_sim}. The contours look gratifyingly similar. As
in the case of the hydro-simulations there is a degeneracy of the
inferred $\sigma_{8}$ with the assumed $\Omega_{\rm m}$ while $X_{J}$
is tightly constrained.

The degeneracy between $\Omega_{\rm m}$ and $\sigma_{8}$ is somewhat
weaker and the constraints on $\sigma_{8}$ (marginalized over the
other two parameter) are somewhat stronger than in the analysis of the
mock spectra.  Note that the constraints on $X_{J}$ are also tighter.
These differences can be traced back to the shape of the thermal cut-off
of the non-linear 1D power spectrum of the gas distribution which is
still affected by the limited dynamical range of the simulations. 

\begin{figure*}
\setlength{\unitlength}{1cm} \centering
\begin{picture}(17,16.5)
\put(-1.5, 6.2){\includegraphics{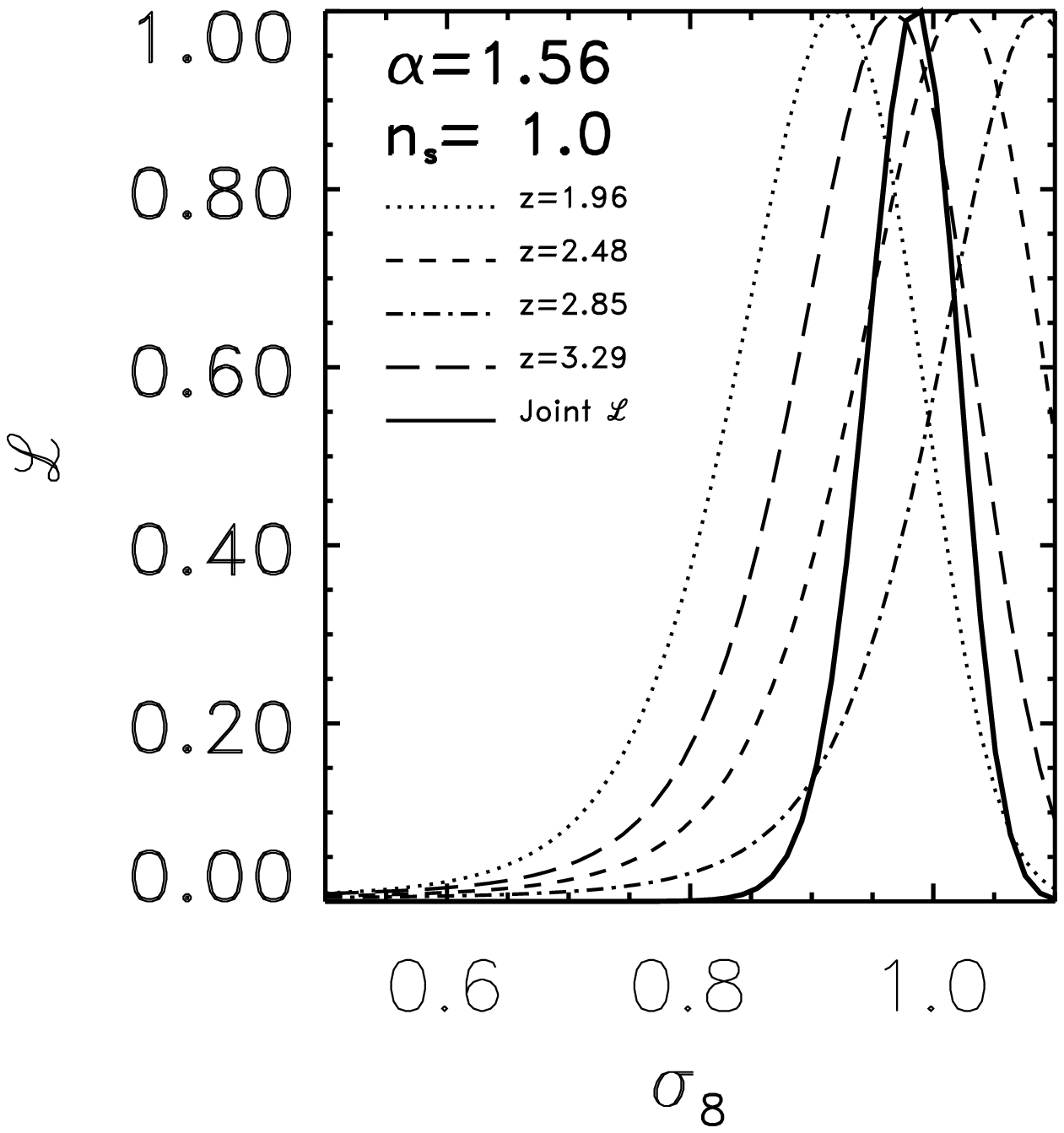}}
\put(8., 6.2){\includegraphics{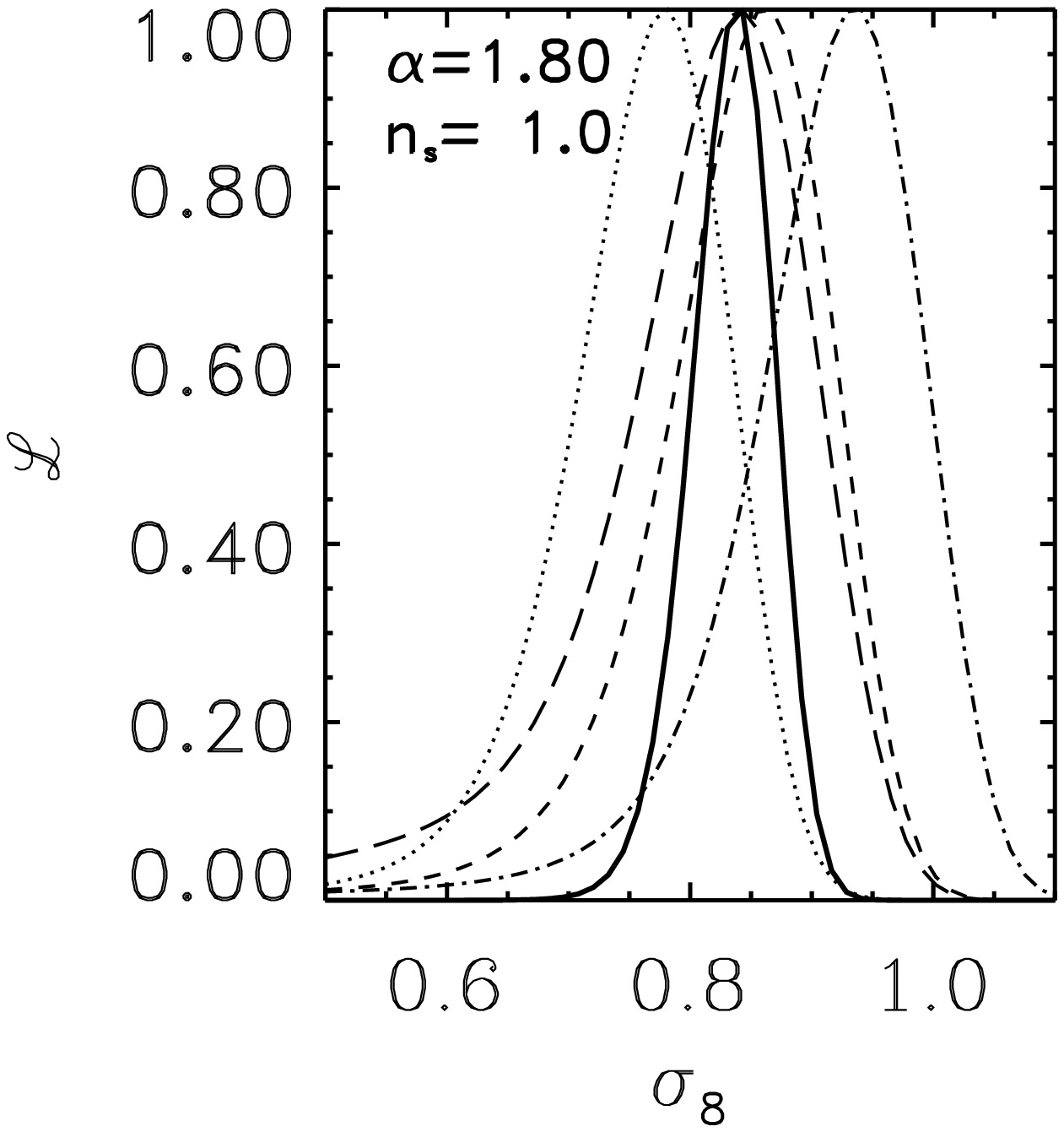}}
\put(-1.5, -2.1){\includegraphics{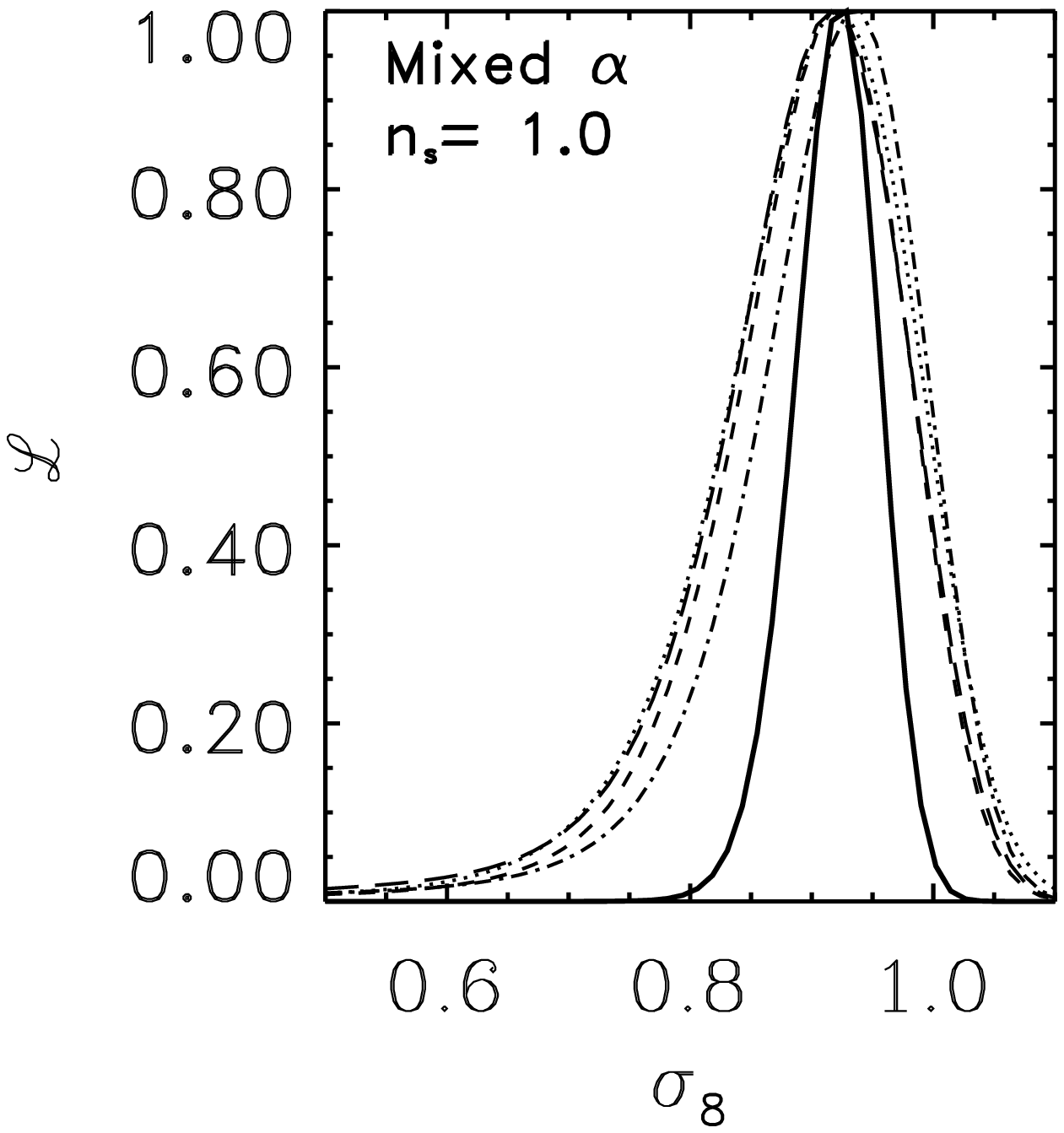}}
\put(8., -2.1){\includegraphics{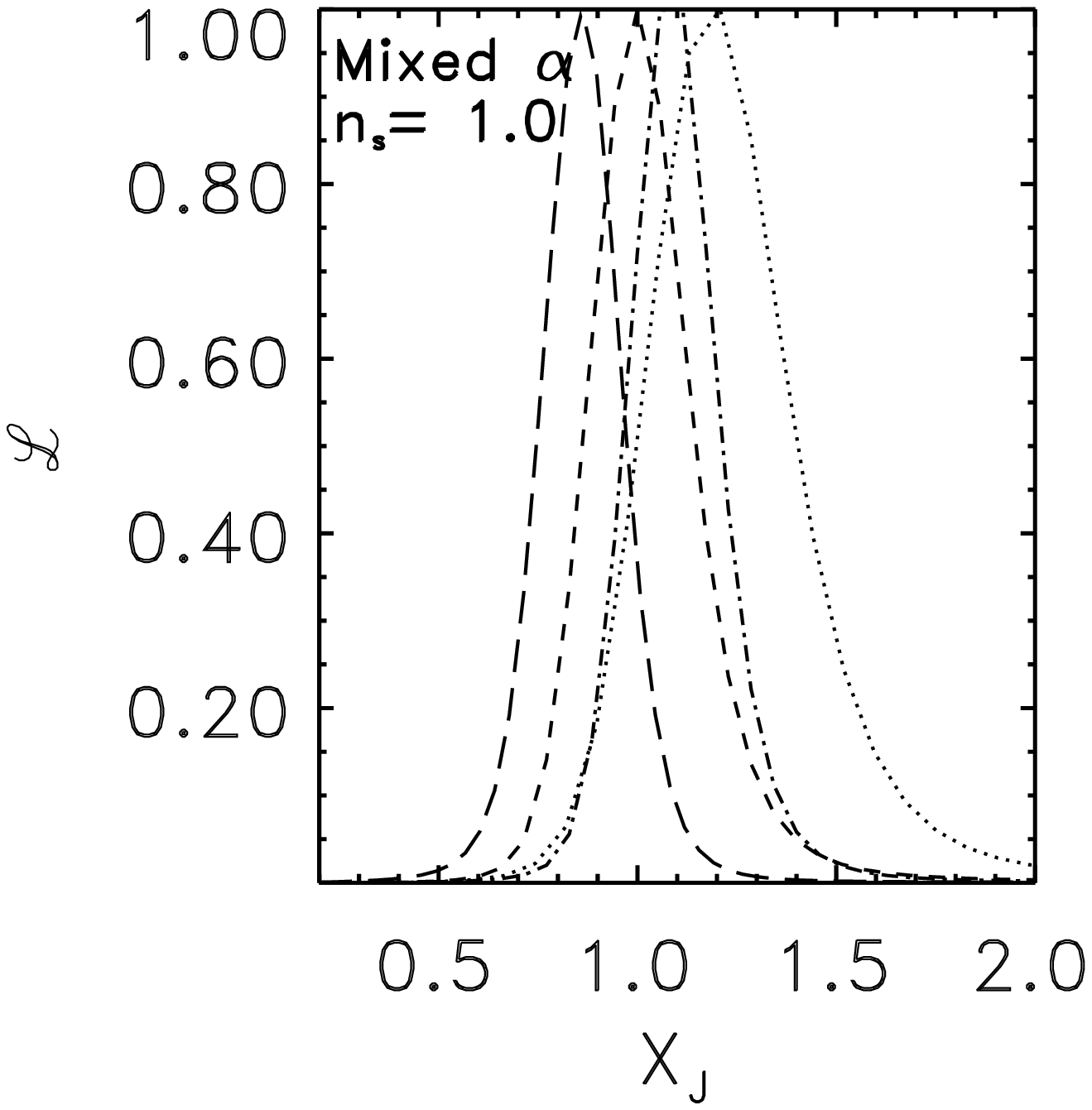}}
\end{picture}
\caption{The  two upper panels show the likelihood distribution for  $\sigma_8$   with
marginalization over $\Omega_{\rm m}$ and $X_J$ for different
redshifts with $\alpha=1.56$ (left panel) and $\alpha = 1.8$ (right
panel). The lower panel show the likelihood distribution for
$\sigma_8$  and $X_J$ for  a model in which $\alpha$ evolves with redshift  from 
$\alpha=1.65$ at $\langle z\rangle=3.29$ rising to $\alpha=1.8$ at $\langle
z\rangle=2.85$ and then $\alpha$ decreasing to $1.65$ and
$1.56$ at $\langle z\rangle =2.48$ and $1.96$ respectively. Such an
evolution would be expected if He became fully reionized at $z\sim
3.2$. The key for the line styles is given in  the upper left panel.}
\label{fig:sgm8}
\end{figure*}

Figure~\ref{fig:sgm8} summarizes the results for the marginalized
likelihood of $\sigma_8$ for all four redshift bins. The upper left
panel shows the likelihood plots for the 4 bins, calculated with
$\alpha=1.56$. The results for the bins with $z=$1.96, 2.48, 2.85, and
3.29 are shown as dotted, dashed, dotted-dashed and long-dashed lines
respectively; the solid curve is the joint likelihood curve.  The
likelihood lines shift slightly between different redshift bins.  The
upper right panel is the same as the upper left panel but for
$\alpha=1.8$. The preferred value of $\sigma_8$ drops here, as
expected, by about 30\%. The lower left panel is the same as the
previous two panels except that $\alpha=$1.56, 1.7, 1.8 and 1.65 for
$z=$1.96, 2.48, 2.85, and 3.29, respectively. Such an evolution is
consistent with reionization of \hep\ at $z\approx3.2$.

The lower right panel of figure~\ref{fig:sgm8} shows the likelihood
curves of the effective Jeans length, $X_J$. Here, we assumed the
mixed $\alpha$ model, \ie\, $\alpha$ values as in the lower left
panel, for each of the four redshift bins.  There is a clear evolution
between the redshift bins with the exception of the $z=2.85$ and $z=2.48$
redshift bins where the evolution is very small. The effective Jeans
length increases with decreasing redshift. This is consistent with an
increased heating rate at redshift $\ga 3.2$ due to helium not yet
being fully ionized.  The increased energy injection during \hep\
reionization should lead to an expansion of the sheets and filaments
responsible for the \lya forest.  Note that the timescale for $X_J$ to
change should be a fair fraction of the Hubble time and that $X_J$ is
therefore not necessarily a good measure of the instantaneous
temperature. We have also examined the effect of changing $n_s$.  Lowering $n_s$
to $0.95$, results in an increase of the estimated power spectrum
amplitude $\sigma_8$ by 4 \%.

In order to show that the models with the preferred parameters of our 
likelihood analysis (discussed later in this section) fit the data well, 
the measured 1D power spectra shown in 
Figure~\ref{fig:PS1D} are compared with the theoretical models.  
The models are shown as the dashed curves and have three $\sigma_8$ values, 0.85,
0.95 and 1.05.  Notice that $\Omega_{\rm m} =0.3$. The values of
$\alpha$ are consistent with the evolution expected for a
reionization of \hep\ at $z\sim 3.2$ (see discussion later in the
section).

\subsection{Constraints on $\alpha$ and $X_J$ with a 
fixed cosmology and normalization}
 
Does the data prefer certain values of $\alpha$? In order to answer
this question more definitely, we carried out an analysis  where
$\alpha$ and $X_J$ are left free while the cosmological
parameters $\Omega_{\rm m}$, $\Omega_\Lambda$ are fixed to $0.3$ and
$0.7$ respectively. The amplitude of power spectrum was 
set to match that of the 2dF galaxy survey normalization (Cole \etal
2005). Since the mean separation between galaxies in the survey is
about $10 \mathrm{h^{-1} Mpc}$, $\sigma_8$ is not directly measured by
the 2dF data. To avoid extrapolating to smaller scales, the
$\sigma_{30}$ amplitude is used. This amplitude is defined as the rms
fluctuations within $30 \mathrm{h^{-1} Mpc}$ spheres, a scale directly
probed by 2dF.  The value of $\sigma_{30}$ measured by 2dF is 0.233
(Percival et al 2002). The mass-galaxy bias ratio is assumed
to be unity and any possible errors of the 2DF measurement (which are
relatively small) is neglected in the likelihood analysis.

\begin{figure*}
\setlength{\unitlength}{1cm} \centering
\begin{picture}(16,20)
\put(-3.,-1){\includegraphics{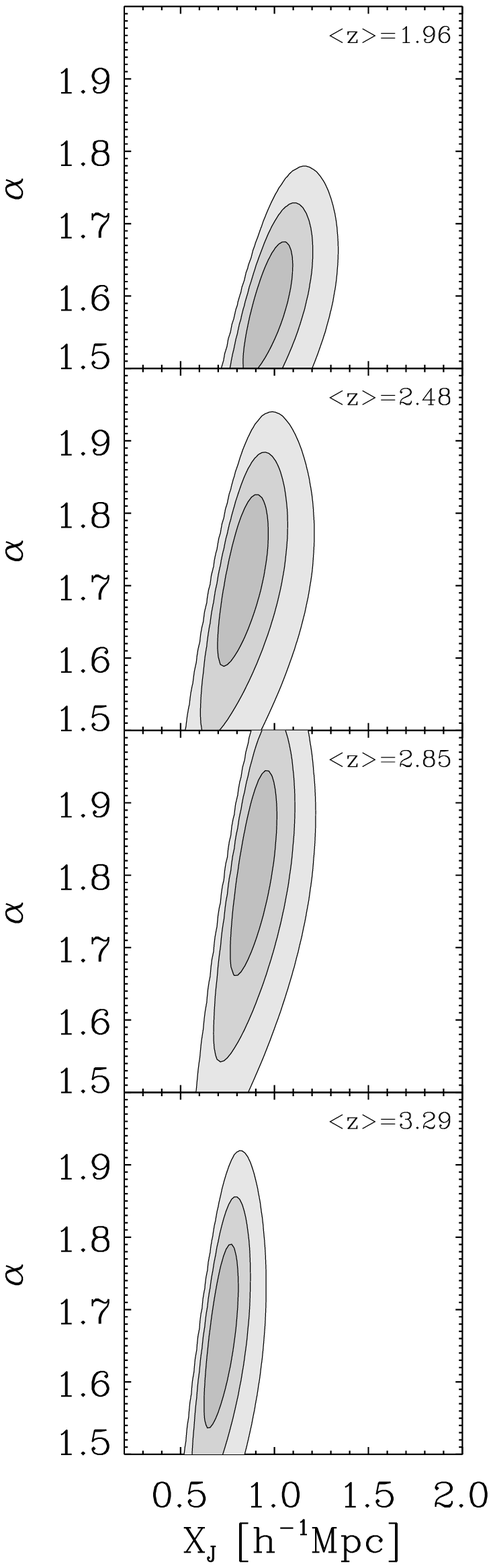}}
\put(3.5,-1){\includegraphics{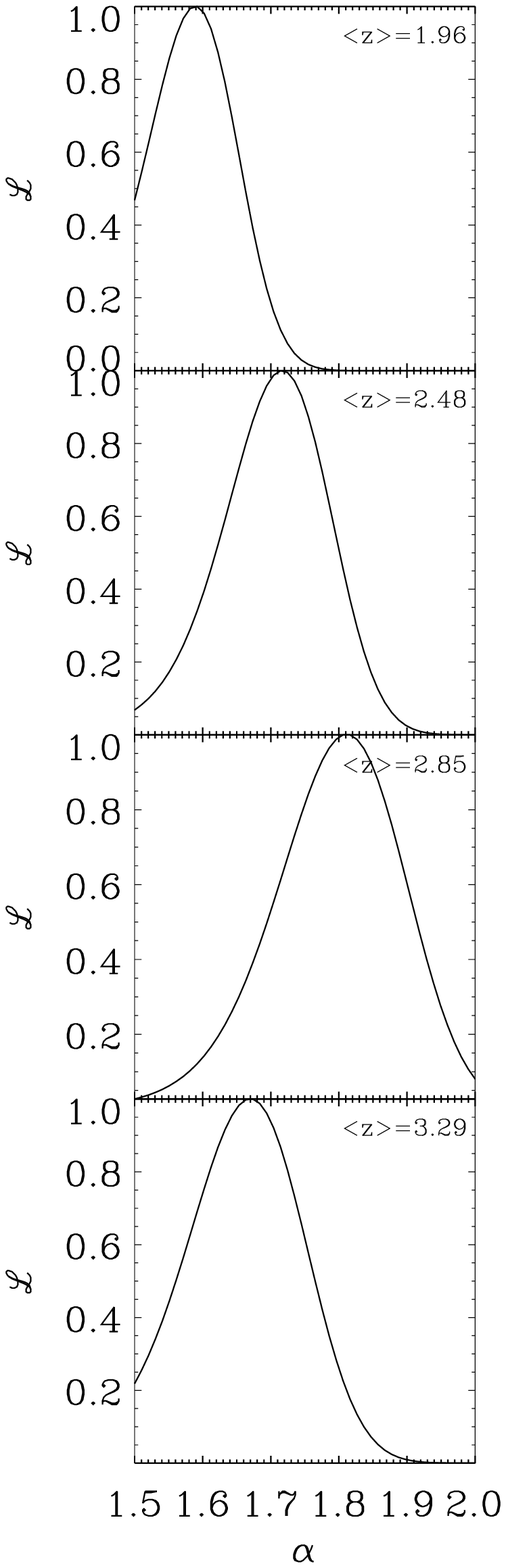}}
\put(9.5,-1){\includegraphics{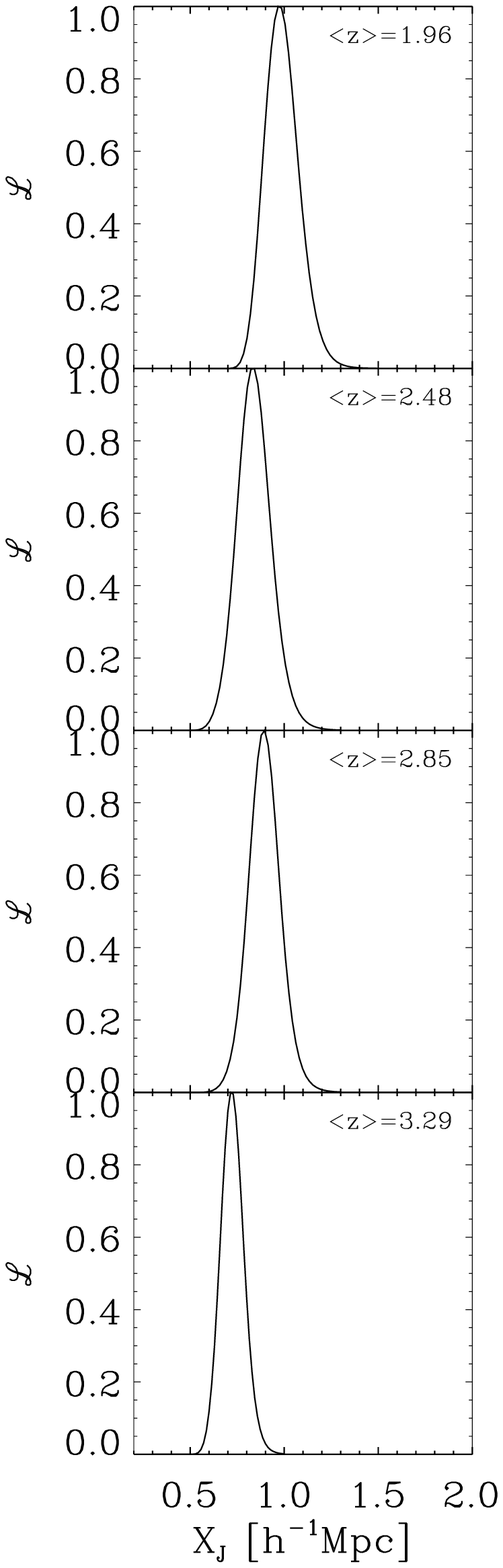}}
\end{picture}
\caption{ Constraints on $\alpha$ and $X_J$ with fixed cosmological
parameters, $\Omega_{\rm m}=0.3$, $\Omega_\Lambda=0.7$ and amplitude
of the matter power spectrum $\sigma_{30}$ as measured by 2dFGRS.}
\label{fig:alpha}
\end{figure*}

Figure~\ref{fig:alpha} shows the likelihood contours for $\alpha$ and
$X_J$ for each redshift bin together with the marginalized
likelihoods. The data prefers values of
$\alpha$ within the assumed physical limits of 1.56 and 2, and appears
to suggest that $\alpha$ evolves with redshift as expected if 
\hep\ reionization indeed occurred at $z\sim 3.2$.  The inferred values
of $X_J$ are consistent with the results obtained in the previous
subsection, and are only weakly dependent on the values of the
cosmological 
parameters.

\begin{figure*}
\setlength{\unitlength}{1cm} \centering
\begin{picture}(14,13)
\put(-0.0, -3.0){\includegraphics{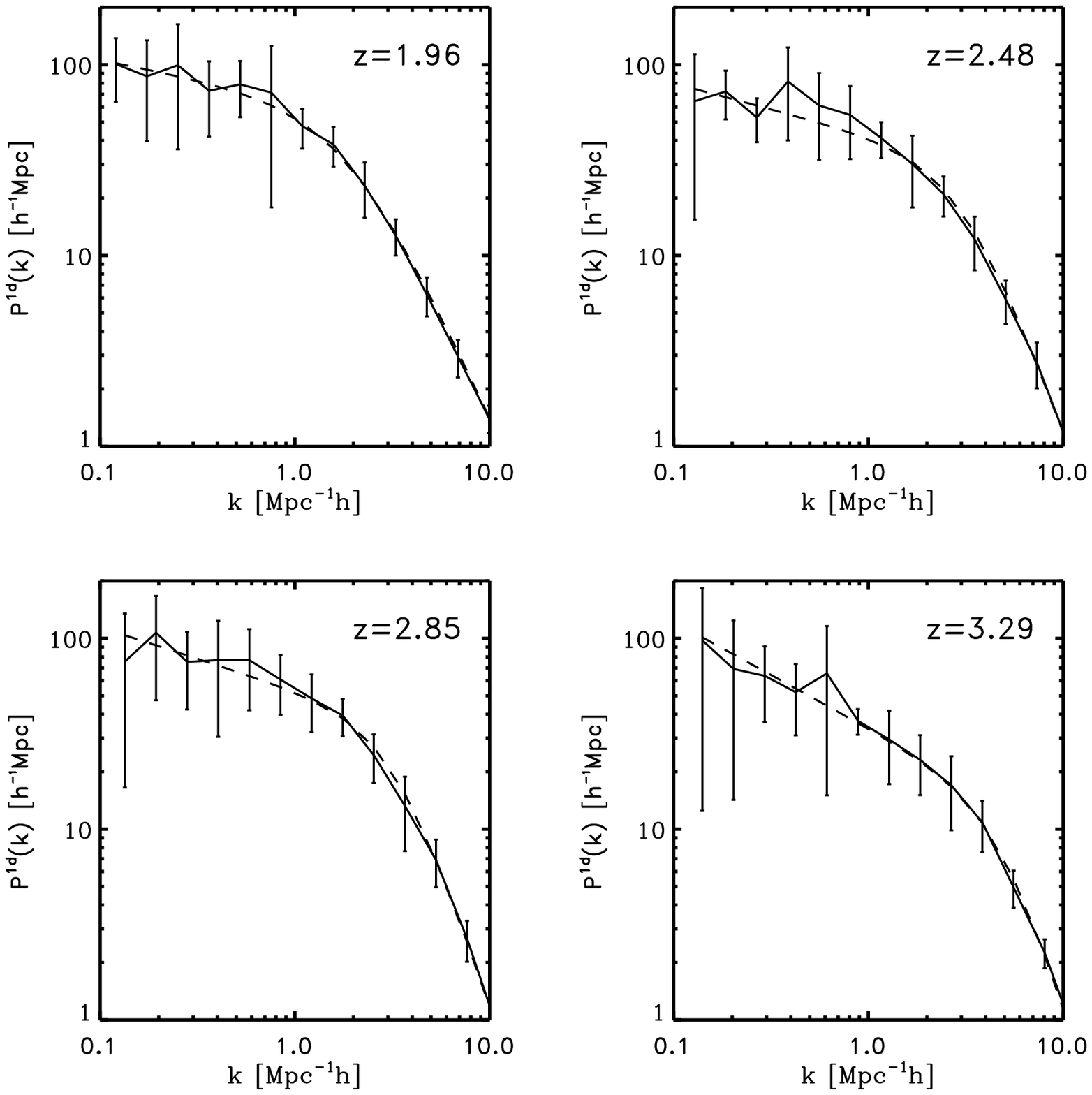}}
\end{picture}
\caption{The measured 1D power spectrum of the gas density in four
redshift bins. The dashed curves show the best-fitting models.}
\label{fig:1dfit}
\end{figure*}

\subsection{Joint constraints of $\Omega_{\rm m}-\sigma_8$ from \lya and WMAP data}

\begin{figure*}
\setlength{\unitlength}{1cm} \centering
\begin{picture}(21,7)
\put(-1.0, -2.2){\includegraphics{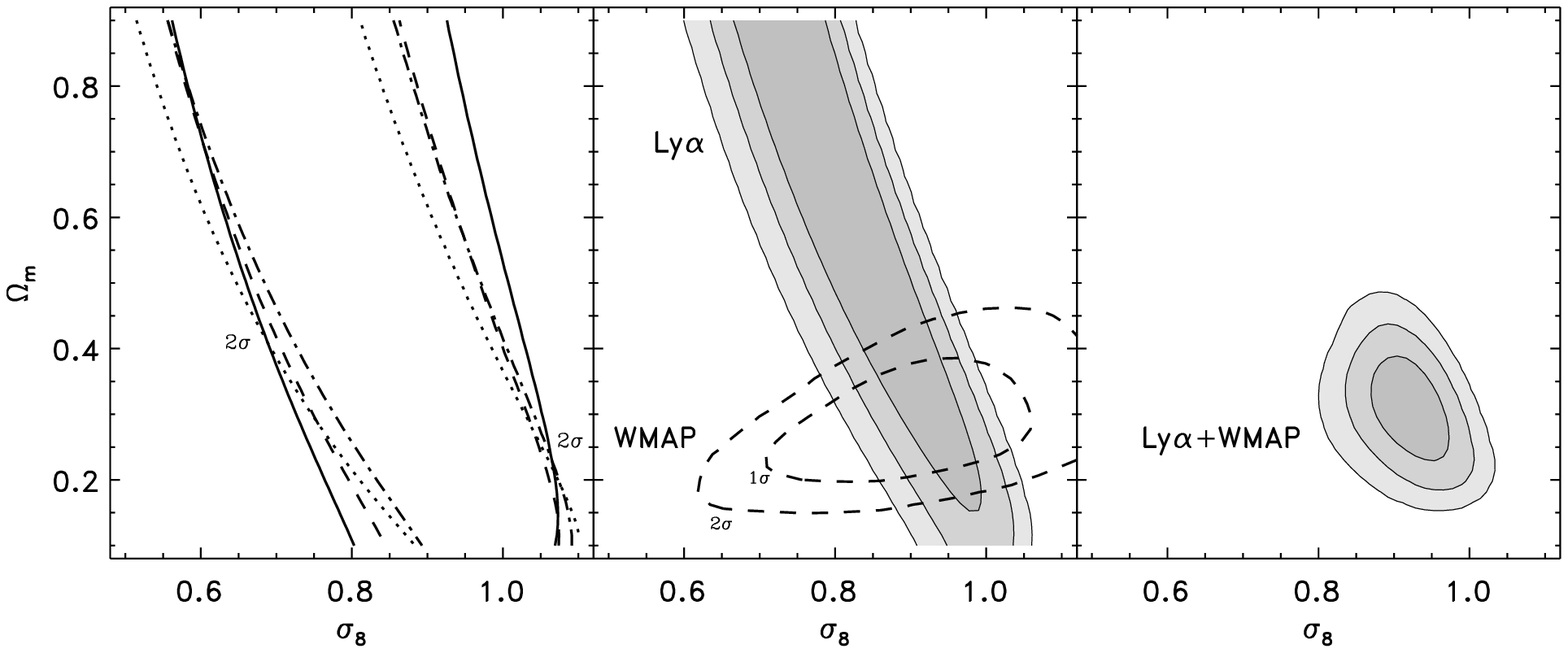}}
\end{picture}
\caption{{\it Left panel:} The likelihood contours in the $\Omega_{\rm
m}-\sigma_8$ plane, after marginalizing over $X_J$ for four different
redshifts for the `mixed' $\alpha$ model discussed in the text.  {\it
Middle panel:} The joint likelihood of the four redshift bins together,
with solid  contours representing the 1,2, and 3-$\sigma$
likelihood.  Also shown are the 1 and 2-$\sigma$ contours (dashed
curves) as obtained from the WMAP data, assuming a Universe with
$\Omega_k$ between -0.3 and 0.3, and a Hubble constant of $72 \pm
8\mathrm{\kms \Mpc^{-1}}$. {\it Right panel}: The joint likelihood of
the \lya data and WMAP data (again assuming the Universe to be flat)
and an HST prior on the Hubble constant.}
\label{fig:fit}
\end{figure*}

As apparent from Figs. 8 and 9 and discussed in section
\ref{cosmopar}, the fluctuation amplitude of the linear matter power
spectrum inferred from the likelihood analysis is somewhat degenerate
with the inferred value for $\Omega_{\rm m}$. This degeneracy is shown
in the left panel of Fig. \ref{fig:fit} which shows the 2-$\sigma$
contours for four redshift bins separately. The likelihood contours of
the four bins are in good agreement.  We have here assumed again the
`mixed' $\alpha$ model for the optical depth density relation and made
the same assumptions as in section \ref{cosmopar}.

As shown in the middle panel of Fig. \ref{fig:fit} this
degeneracy is orthogonal to a similar degeneracy for the CMB
data. The dashed curves show the constraints from the 
WMAP data combined with a prior on the Hubble constant,  
$H_0=72 \pm 8\mathrm{\kms \Mpc^{-1}}$ (Freedman et al. 2001). 
The error contours were calculated using 
COSMOMC (Lewis \& Bridle 2002).  
The  curvature of the Universe, $\Omega_k$, was  assumed to
lie  between $-0.3$ and $+0.3$. 
Without the prior on the Hubble constant  the
contours broaden somewhat but show a similar degeneracy.  
The right panel shows the much tighter constraints on  $\sigma_8$ and 
$\Omega_{\rm m}$ obtained by combining 
our \lya forest data with the WMAP and HST key project data. 
The joint analysis yields the values $\sigma_{8}= 0.9 \pm 0.06$ 
and $\Omega_{\rm m}= 0.3 \pm 0.05$ (in
good agreement with Viel, Haehnelt \& Springel 2004, Seljak et
al. 2005 and Viel \& Haehnelt 2005).
 
\section{The three dimensional power spectrum}
\label{sec:3DPS}

In principle it would be more convenient to infer the 3D matter power
spectrum directly from the data. This would also facilitate a more
direct comparison with the results of Viel et al. (2004) who inferred the 3D
matter power spectrum from the flux power spectrum using an effective
bias method calibrated with numerical simulations.  However, as
discussed above, this requires taking the derivative of  noisy
data. We deal here with this problem by assuming that the 1D power
spectrum is an analytic function. We use a generic curve to fit
the 1D data points. The derivative is then easily obtained. We have 
fitted the following functional form to the measured 1D power-spectrum,
\begin{equation}
f(k; \mathbf{\mu}) = \frac{A_0 k^\gamma}{1+(k/k_0)^\beta} \;\; ,
\label{eq:fit}
\end{equation}
where $\mathbf{\mu} \equiv \left(A_0, \gamma, k_0, \beta\right)$ is
the free parameters vector.

\begin{figure*}
\setlength{\unitlength}{1cm} \centering
\begin{picture}(17,8.5)
\put(-1.5, -2){\includegraphics{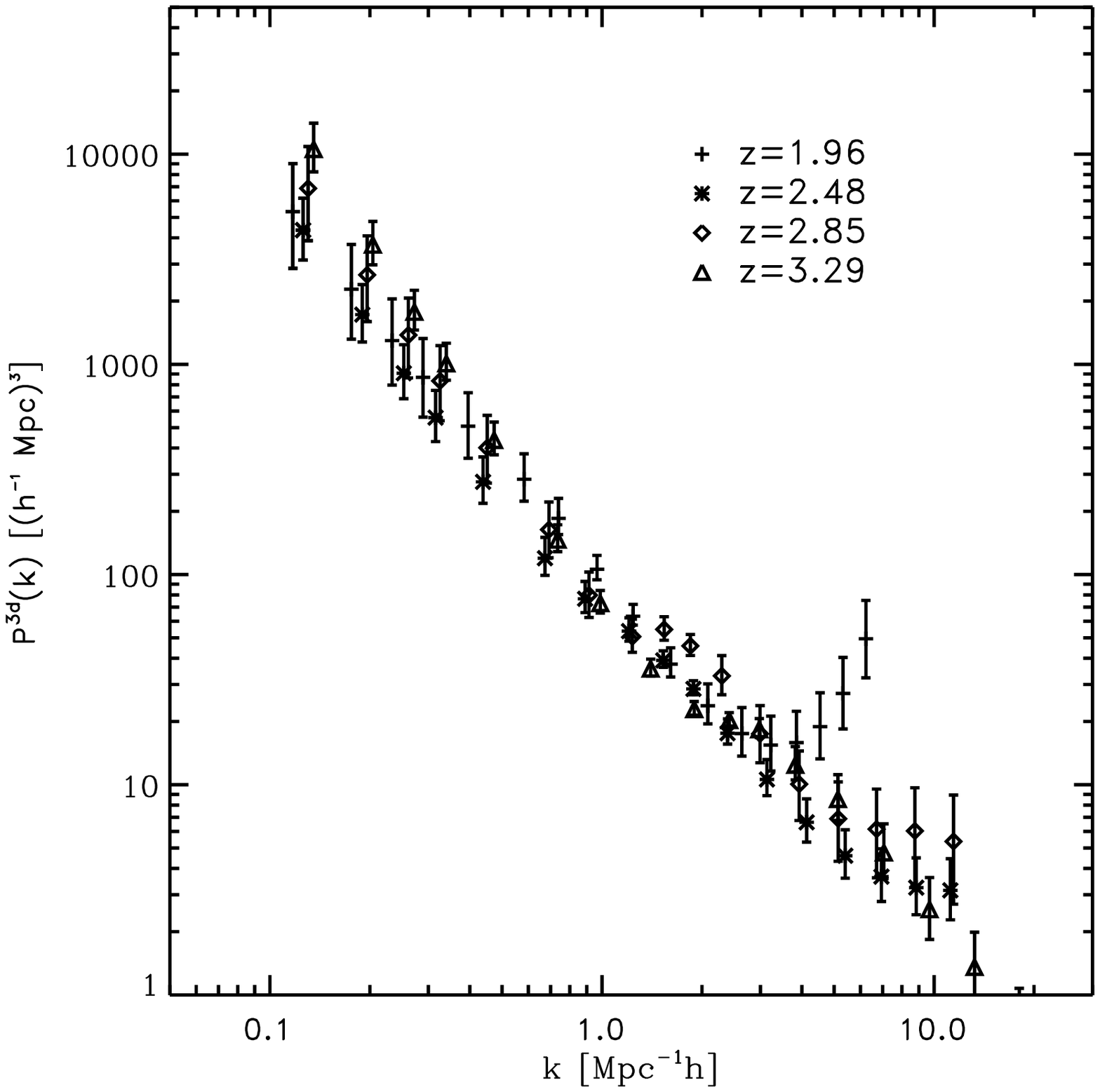}}
\put(8.2, -2){\includegraphics{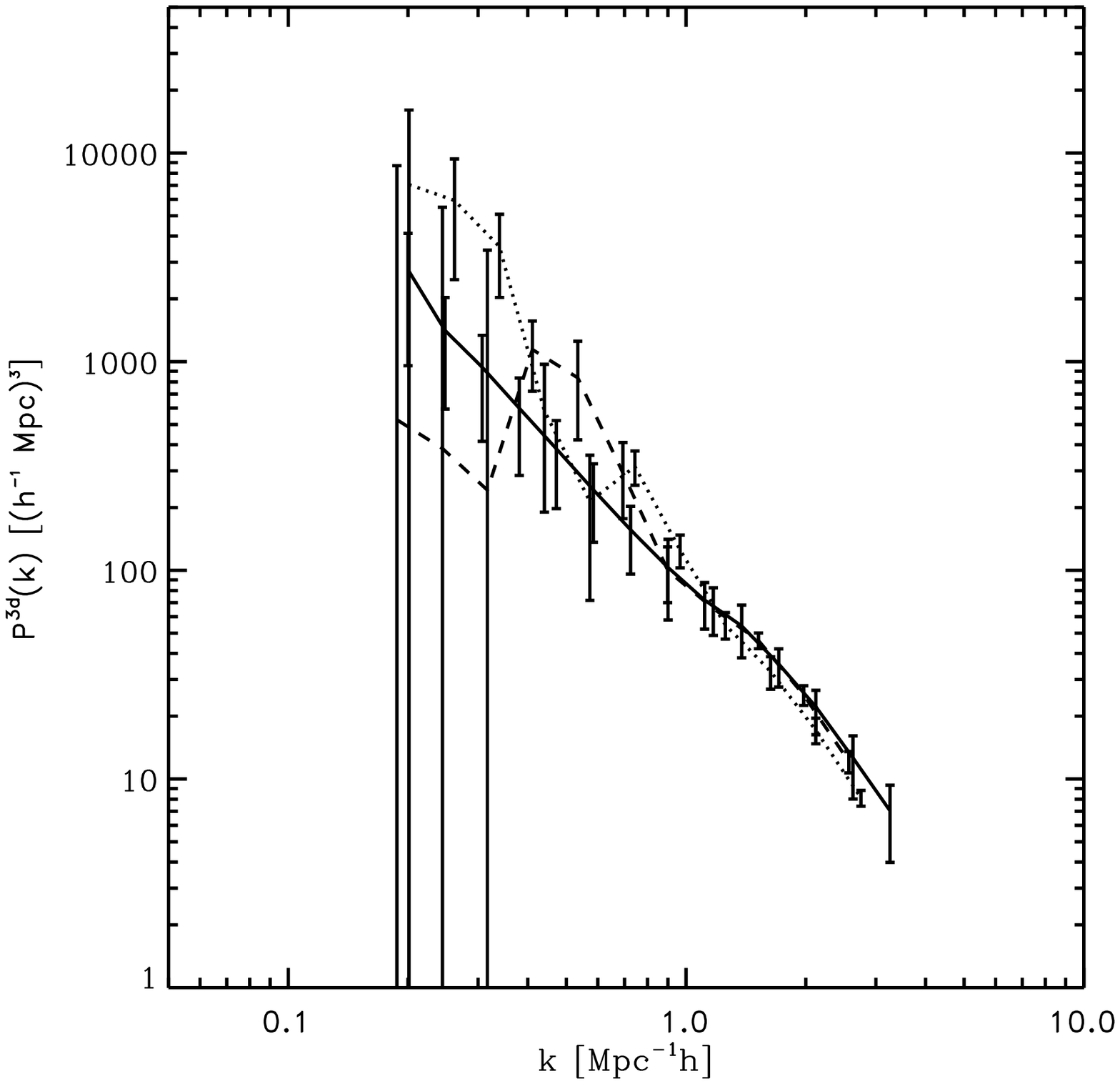}}
\end{picture}
\caption{{\it Left panel:} The linear 3D matter power spectrum at
$z=0$ inferred from observed data at four different redshifts.  The
power spectra have been mapped to $z=0$ assuming a flat Universe with
$\Omega_{\rm m}=0.3$ and $\Omega_\Lambda=0.7$. The power spectra have
been corrected for redshift distortions, Jeans smoothing and for
non-linear effects as described in the text. Notice that the inversion
becomes unstable at scales smaller than the Jeans scale relevant to
the redshift at hand. The right panel shows the mean and error after
combining the results from the four redshifts (solid curve). For
comparison we show the linear matter power spectrum power spectrum
inferred by Viel et al. 2004 from the LUQAS sample (dashed curve) and
from the Croft et al. 2002 sample (dotted curve), as re-analysed by
Viel et al. (2004). }

\label{fig:3Dfit}
\end{figure*}

The main issue that remains is how to assign errors to the derived 3D
power spectrum. In order to estimate the errors, let $\delta
f(k;\mathbf{\mu})$ be the uncertainty in the functional fit due the
errors in the data. Assuming that the fit has the correct functional
form, this error is due to the uncertainty in determining the free
parameters, namely,
\begin{equation}
\delta f = \delta\mathbf{\mu} \cdot \left(\frac{\partial
f}{\partial\mathbf{\mu}}\right)_\mathbf{\mu_0}
\end{equation}
The value of $\delta\mathbf{\mu}$ can easily be estimated from the
fitting procedure used to obtain the most likely values of
$\mathbf{\mu_0}$. Here we chose to do this with  a minimum $\chi^2$
analysis. Once $\delta\mathbf{\mu}$ is known it is straightforward to
show that the uncertainty of the derivative is given by:
\begin{equation}
\delta\left(\frac{\mathrm{d} f}{\mathrm{d} k}\right) = \sum_i
\delta\mu_i \frac{\mathrm{d} \left({\partial f}/{\partial
\mu_i}\right)_\mathbf{\mu_0} }{\mathrm{d} k}.
\label{eq:fit_error}
\end{equation}
The right hand side of equation~\ref{eq:fit_error} is readily calculated
and gives the errors associated with the inverted quantity.

The 1D power spectra are obtained assuming the mixed
$\alpha$ model for the optical depth density relation.  
Figure~\ref{fig:1dfit} shows the functional fit to
the observed 1D power spectra at each redshift bin for the
mixed $\alpha$ model. Equations~\ref{eq:fit} and~\ref{eq:fit_error} are
used to obtain the 3D power spectra. These have then be  scaled to redshift
zero using the linear growth factor and corrections for the
effects of redshift distortions and Jeans smoothing have been
made. For the  latter the most likely Jeans scale found in the
likelihood analysis has been used.

The left panel of figure~\ref{fig:3Dfit} shows the 3D power spectra
for the four redshift bins assuming that $\alpha$ evolves
with redshift as suggested by our likelihood analysis. The figure
clearly shows that the 3D power spectra from the four redshift bins is
consistent down to the scales where the inversion becomes unstable 
and power spectra start to diverge. This
instability is caused by the insufficient information content in the
data below the Jeans smoothing scale. As expected the scale of 
instability varies with redshift. 
Note that the wavenumbers $k$ shown here is the linear $k$, the actual measured $k$ is
larger by roughly a factor of 5 (see Peacock \& Dodds 1991).

The right panel shows the average 3D power spectrum with errors that
reflect the uncertainty in $\alpha$ and the differences between the four
redshift bins. The uncertainty in $\alpha$ is added by taking
the amplitude variation within the range $1.56\le \alpha \le 1.8$.  In
the averaging procedure, the points at scales smaller than the scale of
convergence are ignored. The right panel also shows the 3D
power spectra inferred  by Viel et al. (2004; dashed line) from the
LUQAS sample and from the Croft et al. (2002) sample (dotted line) as
reanalysed by Viel et al. (2004).  There is excellent agreement
between the inferred linear 3D matter power spectrum of this study and
those  obtained by Viel et al. (2004).

This is  gratifying as our inversion method does not require the
assumption of an effective optical depth (see Lidz et al. 2005 who
come to similar conclusions by combining 1 and 2pt statistics of the
flux distribution). Table~\ref{table2} tabulates  our inferred 3D power
spectrum and the $1-\sigma$ errors.

\begin{table}
\label{table2}
\caption{A table showing the 3D power spectrum as estimated from the
direct inversion method. The first column shows the wavenumber $k$. The
second column shows the 3D power spectrum and the third and fourth
columns show the lower and upper error bars around the mean.}
\begin{tabular}{cccc}
\hline
$k$ & $ P^{3D} $ & $ -\Delta P^{3D} $ & $ +\Delta P^{3D} $  \\
$ [\mathrm{Mpc^{-1}h}] $ & $ [\mathrm{h^{-3}Mpc^3}] $ & $ [\mathrm{h^{-3}Mpc^3}] $ & $ [\mathrm{h^{-3} Mpc^3}] $ \\
\hline
     0.200000  &        3023.49  &        1005.55  &        1716.95 \\
     0.248189  &        1325.90  &        577.095  &        760.660 \\
     0.307990  &        877.984  &        372.994  &        495.885 \\
     0.382199  &        586.556  &        210.709  &        296.521 \\
     0.474288  &        410.985  &        129.560  &        187.204 \\
     0.588566  &        263.541  &        84.0443  &        114.476 \\
     0.730379  &        163.626  &        54.6102  &        68.7632 \\
     0.906361  &        116.986  &        41.3544  &        48.0453 \\
      1.12475  &        70.5957  &        24.1203  &        27.1044 \\
      1.39575  &        52.7193  &        16.9332  &        18.2188 \\
      1.73205  &        35.7962  &        9.52174  &        10.8408 \\
      2.14938  &        24.4715  &        3.33336  &        5.43795 \\
      2.66727  &        15.7595  &        2.04052  &        3.73897 \\
      3.30994  &        10.1860  &        1.78284  &        3.32257 \\
      4.10746  &        6.36426  &        5.19802  &        8.69673 \\
 \hline
\end{tabular}
\end{table}

\section{summary \& conclusions}
\label{sec:discuss}

We have presented  a new method for measuring the matter power spectrum
from \lya forest data. The 1D density
field is obtained from the reconstructed line-of-sight optical depth
for \lya absorption, assuming a simple power law relation between 
density and optical depth.  We  thereby follow NH99 \& NH00 and introduce a
cut-off in the optical depth to handle the saturation effects of the
flux in high density regions.
The shape of the non-linear 1D power spectrum of the gas density
does  not depend on the value of the optical depth cut-off.
This allows us to the derive  the amplitude and the shape of the 
1D power spectrum of the gas density separately from 2pt and 1pt
statistics of the flux. The shape of the power spectrum is calculated 
directly form the recovered line-of-sight gas density while the
amplitude of the power spectrum is derived from the width of the PDF
of the gas distribution which we assume to follow a log-normal
distribution. In this way we can measure shape and amplitude of the 
1D power spectrum of the gas density without the need for
determining or assuming a mean flux level, which has proven problematic
in previous measurements of the matter power spectrum utilizing the
flux power spectrum.  Note that the inferred amplitude of the power
spectrum still depends on the assumed power law index $\alpha$, which
describes the relation between the neutral and the total gas density
and depends on the thermal state of the gas.

We have then compared the inferred non-linear 1D power spectrum of the gas density to an
analytical model of the gas power spectrum using a likelihood
analysis.  We have thereby used state-of-the art hydrodynamical
simulations to model the redshift-distortions and the bias between
gas and dark matter distribution and to test our method.  
The \lya-forest data cannot constrain all relevant parameters
simultaneously.  We have therefore restricted our likelihood analysis 
to flat cosmological models.  We only vary a subset of parameters at any given
time setting the remainder to plausible values.  We have also 
performed constraints from a joint analysis of our results 
from the \lya forest data and other data (2dF,CMB,HST).  Finally we
have obtained  a direct estimate of the 3D matter power spectrum from
the 1D gas power spectrum and compare it to previous estimates from
\lya forest data utilizing the flux power spectrum.

Our  main results can be summarized as follows.
\begin{itemize}

\item
By enforcing the cosmological parameters to be the same in all four
redshift bins of our likelihood analysis, we found evidence for
evolution of the density-temperature relation and the thermal
smoothing length of the gas. The inferred evolution is consistent with
that expected if the reionization of \hep\ occurred at $z\sim 3.2$.

\item
Assuming that the Universe is flat and assuming a fixed value for
$n_{\rm s}=1$, we find that the fluctuation amplitude of the matter
power spectrum, $\sigma_8 = 0.9 (\Omega_{\rm m}/0.3)^{-0.3}$ where  the
value of $\alpha$ changes as a function of redshift. 
 The 1$\sigma$ error on $\sigma_8$ at fixed $\Omega_{\rm m}$ and
$\alpha$ is about 10\%.  The thermal smoothing length $X_J$
is also found to be tightly constrained (to within 10 percent).

\item
A joint analysis of the \lya forest and the WMAP CMB data together with 
a prior on the Hubble constant, yields tight constraints on the
fluctuation amplitude and the matter density,
$\sigma_{8}= 0.9 \pm 0.09 $, $\Omega_{\rm m}=0.30 \pm 0.05 $.

\item
The inferred linear 3D matter power spectrum agrees well with that
obtained by Viel et al. 2004  with a very
different analysis technique.
\end{itemize} 

The good agreement of our  results for the amplitude and shape of the
matter power spectrum with those of previous studies of  \lya forest data is very
reassuring, as the systematic  uncertainties differ significantly 
for the different methods employed by these studies. The  independent
constraints on the thermal state of the gas suggest that the inferred 
peak in the photo-heating rate of helium at $z\sim 3$ has affected the 
flux distribution of the \lya forest in a measurable way.

\section*{Acknowledgements}
The simulations were run on the COSMOS (SGI Altix 3700) supercomputer
at the Department of Applied Mathematics and Theoretical Physics in
Cambridge and on the Sun Linux cluster at the Institute of Astronomy
in Cambridge. COSMOS is a UK-CCC facility which is supported by HEFCE
and PPARC.  The authors acknowledge support from the European
Community Research and Training Network ``The Physics of the
Intergalactic Medium''.  MV and SZ would like to acknowledge the
hospitality of the Institute of Theoretical Physics at the Technion,
Haifa.  AN and SZ thank the IoA, Cambridge for hospitality. 

{}
\end{document}